\documentclass[11pt, draftclsnofoot, onecolumn]{IEEEtran}
\usepackage[caption=false,font=footnotesize]{subfig}
\usepackage{graphicx}
\usepackage{booktabs}
\usepackage{amsmath}
\usepackage{amssymb}
\usepackage{stfloats}
\usepackage{bm} 
\usepackage{color, cite}
\usepackage{algorithm}
\graphicspath{{figures/}}

\usepackage{enumerate,amsthm,}

\theoremstyle{plain}

\newtheorem{definition}{Definition}

\newtheorem{theorem}{Theorem}
\newtheorem{lemma}{Lemma}
\newtheorem{corollary}{Corollary}

\newtheorem{proposition}{Proposition}

\bstctlcite{BSTcontrol}

\begin{document}

\title{Multi-dimensional Spectral Super-Resolution with Prior Knowledge via Frequency-Selective Vandermonde Decomposition and ADMM}
\author{
Yinchuan~Li, Xiaodong Wang, \emph{Fellow}, \emph{IEEE}, Zegang~Ding, \emph{Member}, \emph{IEEE}
\thanks{
Y. Li is with the School of Information and Electronics, Beijing Institute of Technology, Beijing 100081, China, and the Electrical Engineering Department, Columbia University, New York, NY 10027, USA (e-mail: yinchuan.li.cn@gmail.com).

X. Wang is with the Electrical Engineering Department, Columbia University, New York, NY 10027, USA (e-mail: wangx@ee.columbia.edu).

Z. Ding is with the School of Information and Electronics, Beijing Institute of Technology, Beijing 100081, China, and the Beijing Key Laboratory of Embedded Real-time Information Processing Technology, Beijing 100081, China (e-mail: z.ding@bit.edu.cn).
}

}
\maketitle

\begin{abstract}

This paper is concerned with estimation of multiple frequencies from incomplete and/or noisy samples based on a low-CP-rank tensor data model where each CP vector is an array response vector of one frequency. Suppose that it is known a priori that the frequencies lie in some given intervals, we develop efficient super-resolution estimators by exploiting such prior knowledge based on frequency-selective (FS) atomic norm minimization. We study the MD Vandermonde decomposition of block Toeplitz matrices in which the frequencies are restricted to lie in given intervals. We then propose to solve the FS atomic norm minimization problems for the low-rank spectral tensor recovery by converting them into semidefinite programs based on the MD Vandermonde decomposition. We also develop fast solvers for solving these semidefinite programs via the alternating direction method of multipliers (ADMM), where each iteration involves a number of refinement steps to utilize the prior knowledge. Extensive simulation results are presented to illustrate the high performance of the proposed methods.

\end{abstract}

\section{Introduction}

Multi-dimensional (MD) arrays, i.e., tensors, arises naturally in many applications~\cite{ghadermarzy2017near,gandy2011tensor,wei2018tensor,ashraphijuo2017characterization,ashraphijuo2017rank}, including radar and communications signal processing~\cite{nion2010tensor,de2007parafac}, visual data reconstruction~\cite{liu2012tensor}, data mining~\cite{acar2005modeling}, seismic data processing~\cite{ely20135d}. In many areas of signal processing applications, the tensor is a superposition of a small number of MD complex sinusoids, and it is desired to infer the MD spectral contents of a measured tensor. In particular, in MIMO radar and communication systems~\cite{skolnik1970radar,rappaport1996wireless}, where the signal propagation from a transmitter to a receiver can be modelled by an MD tensor, with each dimension representing time delays, Doppler shifts, directions of arrival/departure~\cite{berger2010signal,nion2010tensor,zheng2017super,tse2005fundamentals}. Estimating these parameters involves the spectral tensor recovery and frequency super-resolution, which is crucial for target localization and tracking as well as coherent detection~\cite{li2019multi,tse2005fundamentals}.

Many existing works on one-dimensional (1D) spectral super-resolution have been developed based on the compressed sensing (CS)~\cite{donoho2006compressed,candes2011compressed,zhang2018recovery} theory. Since the frequencies are continuous parameters, conventional CS tools suffer performance degradation when discretizing the signal by a finite discrete dictionary~\cite{stankovic2013compressive,jokanovic2015reduced,studer2012recovery}. Continuous sparse recovery methods, e.g., the atomic norm (AN)~\cite{tang2013compressed,yang2014exact,bhaskar2013atomic} minimization techniques, have attracted considerable interest in spectral super-resolution. In~\cite{tang2013compressed}, the atomic norm approach was proposed for 1D spectral estimation, which can recover the frequency under sub-Nyquist sampling, and it was extended to higher dimensions in~\cite{chi2015compressive,chen2014robust,xu2014precise}. In~\cite{zheng2017super,li2019multi,tan2014direction,tsai2018millimeter}, methods based on atomic norm for delay-Doppler and direction-of-arrival estimation in radar systems and for the channel estimation in wireless communications were developed. In~\cite{yang2016vandermonde2}, the Vandermonde decomposition of multi-level (ML) Toeplitz
matrices was investigated, which can be used for low-rank spectral tensor recovery and MD frequency super-resolution.

The above-mentioned works assume no prior knowledge on the frequencies to be estimated. On the other hand, certain prior knowledge is available in many applications. For instance, in radar systems, one can set up a surveillance area where a target may appear, and the speed range of a particular target may be known. In communication systems, the range of signal delays and directions of arrival/departure maybe known~\cite{beygi2015multi}. In a rotating mechanical system, the frequencies of the supply lines or interfering harmonics may be known~\cite{wirfalt2011subspace}. Moreover, in a precipitation radar, it is possible to know the spectrum widths of echoes from certain weather phenomena based on previous observations~\cite{doviak2014doppler}. Hence, using such prior knowledge to improve the performance of spectral estimation has attracted considerable interest~\cite{linebarger1995incorporating,zachariah2013line}.

In~\cite{mishra2015spectral}, a constrained atomic norm formulation for 1D frequency estimation was proposed based on the theory of positive trigonometric polynomials~\cite{dumitrescu2007positive}. In~\cite{yang2018frequency}, the frequency-selective (FS) Vandermonde decomposition of positive semidefinite Toeplitz matrices was proposed, which can restrict the frequency to lie in a given interval. Based on the FS Vandermonde decomposition, the FS atomic norm minimization problem for 1D frequency estimation was converted into a semidefinite programming (SDP) formulation. In~\cite{chao2016extensions}, an SDP formulation of the FS atomic norm is obtained by using a technique different from the FS Vandermonde decomposition. In~\cite{yang2016weighted}, a weighted atomic norm approach was proposed as an approximate but faster implementation of the FS atomic norm. Unfortunately, the above methods all focus on 1D frequency super-resolution problems, and they are not straightforward to be extended to higher dimensional problems due to the fundamental difficulty of generalizing the classical Caratheodory's theorem~\cite{caratheodory1911zusammenhang} to higher dimensions. Moreover, the computational load of those SDP problems is very high since they involve additional constraints to incorporate the prior knowledge, making them difficult to be implemented in practice.

In this paper, the $d$-dimensional ($d\geq2$) FS Vandermonde decomposition of $d$-level ($d\geq2$) block Toeplitz matrices for low-rank spectral tensor recovery and super-resolution is studied. Assume that the frequencies lie in certain given intervals, we solve the MD-FS atomic norm minimization problems by converting them into SDP formulations based on the MD-FS Vandermonde decomposition. We show that the equivalence between the MD-FS atomic norm minimizations and our proposed SDP formulations is guaranteed under the condition that the MD spectral tensor is low rank. Moreover, the dual problems of the original optimization problems are given, where the dual polynomials can be used for the MD spectral super-resolution. Since the proposed SDPs contain many constraints to utilize the prior knowledge, solving them involves high complexity. We therefore design iterative algorithms based on the alternating direction method of multipliers (ADMM)~\cite{boyd2011distributed} to obtain approximate solutions, where each iteration involves refinement steps to handle the problem that multiple SDP conditions should be simultaneously satisfied. Extensive numerical simulation results are provided to illustrate the performance advantage of the proposed methods over traditional approaches.

The remainder of the paper is organized as follows. In Section II, we present the preliminaries and set up the problems of low-rank spectral tensor recovery with prior knowledge. In Section III, we present the MD-FS atomic norm minimization problems for spectral tensor recovery with prior knowledge and their dual problems for MD spectral super-resolution. Then, we present the MD-FS Vandermonde decomposition results and convert the MD-FS atomic norm minimization problems into SDP formulations. In Section IV, we develop fast solvers for low-rank spectral tensor recovery with prior knowledge based on ADMM. Simulation results are presented in Section V. Section VI concludes the paper.

\section{Preliminaries \& Problem Formulation}

\subsection{Notations and Preliminaries}
Notations used in this paper are as follows. Boldface letters denote vectors and matrices. Uppercase calligraphic letters denote tensors. $\mathbb{R}$ and $\mathbb{C}$ denote the sets of real and complex numbers, respectively. $\Re$ and $\Im$ return the real and imaginary parts of a complex argument, respectively. $(\cdot)^*$, $(\cdot)^T$, $(\cdot)^H$, ${\rm rank}(\cdot)$ and ${\rm Tr}(\cdot)$ denote the conjugate, transpose, conjugate transpose, rank and trace operators, respectively. ${\rm sign}(\cdot)$ denotes the sign function. ${\rm diag}(\cdot)$ denotes the diagonal matrix whose diagonal entries are the input vector. $|\cdot|$ denotes the magnitude of a scalar. $\|\cdot\|_2$ and $\|\cdot\|_F$ respectively denote the $\ell_2$ and Frobenius norms. In particular, for a tensor ${\mathcal{X}}\in \mathbb{C}^{N_1 \times N_2 \times \cdots \times N_d}$, the Frobenius norm is defined as $\|\mathcal{X}\|_{F} = \langle \mathcal{X}, \mathcal{X} \rangle = \sqrt{\sum_{n_1=1}^{N_1}\cdots \sum_{n_d=1}^{N_d} |\mathcal{X}(n_1,...,n_d)|^2}$, and the inner product is defined as $\langle \mathcal{X}, \mathcal{Y} \rangle =  \langle {\rm vec}(\mathcal{X}), {\rm vec}(\mathcal{Y}) \rangle = ({\rm vec}(\mathcal{Y}))^H {\rm vec}(\mathcal{X})$ with ${\rm vec}(\cdot)$ denoting vectorization operator.

{\bf{Hermitian trigonometric polynomial:}} 
$d$ Hermitian trigonometric polynomials of degree one are given by
\begin{align}
\label{eq:htps}
g_i(x) =&~ r_{1,i} x^{-1} + r_{0,i} + r_{-1,i} x,~i=1,...,d,
\end{align}
with $r_{-1,i} = r_{1,i}^*, ~r_{0,i} \in \mathbb{R}$. When $x = e^{i2\pi f}$ with $f \in {[0,1)}$, we define $g_i(f) \triangleq g_i(e^{i2\pi f})$, i.e.,
\begin{align}
g_i(f) = r_{1,i} e^{-i2\pi f} + r_{0,i} + r_{1,i}^*e^{i2\pi f} = r_{0,i} + 2\Re\{r_{1,i}e^{-i2\pi f}\},~i=1,...,d.
\end{align}

{\bf{Block Toeplitz matrix:}}
An ${N_1N_2 \times N_1N_2}$ block Toeplitz matrix $\bm T \triangleq \bm T(\bm B)$ is formed by taking as input a ${(2N_1-1)\times (2N_2-1)}$ matrix
\begin{eqnarray}
\bm B = [\bm b_{-N_2+1},\bm b_{-N_2+2},...,\bm b_{N_2-1}],
\end{eqnarray}
with $\bm b_{j} =  [b_{j}^{-N_1+1},b_{j}^{-N_1+2},...,b_{j}^{N_1-1}]^T,~j = -N_2+1,...,N_2-1$,
and outputing
\begin{eqnarray}
\label{eq:block-Toeplitz}
{\bm{T}}(\bm B) = \left[ {\begin{array}{*{20}{c}}
	{{\rm Toep}(\bm b_{0})} & {{\rm Toep}(\bm b_{-1})} & \cdots & {{\rm Toep}(\bm b_{1-N_2})}\\
	{{\rm Toep}(\bm b_{1})} & {{\rm Toep}(\bm b_{0})} & \cdots & {{\rm Toep}(\bm b_{2-N_2})}\\
	\vdots & \vdots & \ddots & \vdots \\
	{{\rm Toep}(\bm b_{N_2-1})} & {{\rm Toep}(\bm b_{N_2-2})} & \cdots & {{\rm Toep}(\bm b_{0})}
	\end{array}} \right],
\end{eqnarray}
where ${\rm Toep}(\cdot)$ denotes the Toeplitz matrix whose first column is the last $N_1$ elements of the input vector, i.e.,
\begin{align}
\label{eq:Toeplitz}
{{\rm Toep}(\bm b_{j})} =
\left[ {\begin{array}{*{20}{c}}
	{b_{j}^{0}} & {b_{j}^{-1}} & \cdots & {b_{j}^{-N_1+1}}\\
	{b_{j}^{1}} & {b_{j}^{0}} & \cdots & {b_{j}^{-N_1+2}}\\
	\vdots & \vdots & \ddots & \vdots \\
	{b_{j}^{N_1-1}} & {b_{j}^{N_1-2}} & \cdots & {b_{j}^{0}}
	\end{array}} \right] \in \mathbb{C}^{N_1 \times N_1}. 
\end{align}

{\bf{ML block Toeplitz matrix:}}
For a $d$-way tensor ${\cal B}^d \in \mathbb{C}^{(2N_1-1)\times \cdots \times(2N_d-1)}$, define $N_D \triangleq \prod_{i=1}^d N_i$, then a $d$-level block Toeplitz matrix $\bm T^{d} \triangleq \bm T^{d}({\cal B}^d) \in \mathbb{C}^{N_D \times N_D}$ is defined by taking ${\cal B}^d$ as the input and outputing recursively as
\begin{eqnarray}
\label{eq:level-Toeplitz}
{\bm T^{i}}({\cal B}^i) = \left[ {\begin{array}{*{20}{c}}
	{\bm T^{i-1}}({\cal B}^{i-1}(0)) &  \cdots & {\bm T^{i-1}}({\cal B}^{i-1}(1-N_d))\\
	{\bm T^{i-1}}({\cal B}^{i-1}(1)) &  \cdots & {\bm T^{i-1}}({\cal B}^{i-1}(2-N_d))\\
	\vdots &  \ddots & \vdots \\
	{\bm T^{i-1}}({\cal B}^{i-1}(N_d-1)) &  \cdots & {\bm T^{i-1}}({\cal B}^{i-1}(0))
	\end{array}} \right],~i=1,...,d,
\end{eqnarray}
where ${\cal B}^{i-1}(j) = {\cal B}^{i}(:,...,:,j)$. And for $i=1$ and $i=2$, \eqref{eq:level-Toeplitz} becomes \eqref{eq:Toeplitz} and \eqref{eq:block-Toeplitz}, respectively. If we decompose ${\bm T^{d}}$ into $d$-level blocks, and denote ${\bm T^{d}}(...;m_i,n_i;...)$ as the $(m_i,n_i)$-th block at the $i$-th level of $\bm T^{d}$, i.e., ${\bm T^{d}}(m_1,n_1;m_2,n_2;...;m_d,n_d)$ denotes the $((m_1-1)\prod_{i=2}^d N_i + (m_2-1)\prod_{i=3}^d N_i + ... + m_d, (n_1-1)\prod_{i=2}^d N_i + (n_2-1)\prod_{i=3}^d N_i + ... + n_d )$-th element in ${\bm T^{d}}$, then we can write
\begin{align}
{\bm T^{d}}(m_1,n_1;m_2,n_2;...;m_d,n_d) = {\cal B}^d(m_1-n_1,m_2-n_2,...,m_d-n_d),\\
~m_1,n_1 = 1,...,N_1;...;m_d,n_d=1,...,N_d. \nonumber
\end{align}

{\bf{ML block Toeplitz matrix \& MD complex sinusoid:}} Define $\bm s(f,N) \triangleq [1,e^{i2\pi f},...,e^{i2\pi(N-1) f}]^T\in \mathbb{C}^{N \times 1}$ and 
\begin{align}
\label{MDsimusoid}
{\cal A}(\bm f) = \bm s(f_1,N_1) \otimes \bm s(f_2,N_2) \otimes \cdots \otimes \bm s(f_d,N_d),
\end{align}
where $\bm f = [f_1,...,f_d]^T$ and $\otimes$ is the tensor/outer product defined via $[\mathbf{u} \otimes \mathbf{v}]_{i, j}=u_{i} v_{j}$. Define 
\begin{align}
\bm{\bar{s}}(f,N) \triangleq [e^{i2\pi(1-N) f},...,e^{i2\pi(-1) f},1,e^{i2\pi f},...,e^{i2\pi(N-1) f}]^T\in \mathbb{C}^{(2N-1) \times 1},
\end{align}
then we have for an input
\begin{align}
{\cal B}^d = \bm{\bar{s}}(f_1,N_1) \otimes \bm{\bar{s}}(f_2,N_2) \otimes \cdots \otimes \bm{\bar{s}}(f_d,N_d),
\end{align}
a rank-1 ML block Toeplitz matrix has the form 
\begin{align}
{\bm T^{d}}(m_1,n_1;m_2,n_2;...;m_d,n_d) = e^{i 2 \pi (m_1-n_1) f_{1}} e^{i 2 \pi (m_2-n_2) f_{2}} \cdots e^{i 2 \pi (m_d-n_d)f_{d}}, \\
~m_1,n_1 = 1,...,N_1;...;m_d,n_d=1,...,N_d, \nonumber 
\end{align}
and hence 
\begin{align}
\label{eq:Td-vecS}
{\bm T^{d}} = {\rm vec}({\cal A}(\bm f)) ({\rm vec}({\cal A}(\bm f)))^H.
\end{align}

{\bf{ML block Toeplitz matrix \& trigonometric polynomials:}}
Given $d$ trigonometric polynomials $g_1,...,g_d$ as in \eqref{eq:htps}, denoting $\bar{N}_{D} \triangleq \prod_{i=1}^d (N_i-1)$, we define a $d$-level block Toeplitz matrix $\bm T^d_{g_i} \in \mathbb{C}^{\bar{N}_{D} \times \bar{N}_{D}}$ with respect to $g_i$ for $i = 1,...,d$ as
\begin{align}
\label{dTgd}
{\bm T^{d}_{g_i}}(m_1,n_1;m_2,n_2;...;m_d,n_d) = \sum_{k=-1}^{1} r_{k,i} {\cal B}^d(m_1-n_1,...,m_i-n_i-k,...,m_d-n_d), \\
~m_1,n_1 = 1,...,N_1-1;...;m_d,n_d=1,...,N_d-1. \nonumber
\end{align}

\subsection{Problem Formulation}

Consider a $d$-way ($d\geq 2$) tensor ${\mathcal{X}}\in \mathbb{C}^{N_1 \times N_2 \times \cdots \times N_d}$, where each entry can be expressed as a superposition of $r$ $d$-dimensional complex sinusoids 
\begin{align}
\label{eq:tensorX}
{\mathcal{X}}(k_1,...,k_d) = \sum_{\ell=1}^{r} \sigma_{\ell} e^{i2\pi k_1 f_{1,\ell}} e^{i2\pi k_2 f_{2,\ell}} \cdots e^{i2\pi k_d f_{d,\ell}},
\end{align}
where $k_i = 0,...,N_i-1,~i=1,...,d$, $f_{i,\ell} \in {[0,1)}$ and $\sigma_{\ell}$ are the frequencies and the complex gain associated with each $1\leq \ell \leq r$, respectively.

Denote $\bm a_i^{\ell} \triangleq \bm s(f_{i,\ell},N_{i}),~i = 1,...,d, ~\ell = 1,...,r$, then \eqref{eq:tensorX} becomes the following CP decomposition 
\begin{align}
\label{eq:tensorX2}
{\mathcal{X}} = \sum_{\ell=1}^{r} \sigma_{\ell} \bm a_1^{\ell} \otimes \bm a_2^{\ell} \otimes \cdots  \otimes \bm a_d^{\ell}.
\end{align}
Assume that the measurement data model follows
\begin{align}
\label{eq:y-1}
{\mathcal{Y}} = {\mathcal{P}} \odot {\mathcal{X}} + {\mathcal{N}},
\end{align}
where $\odot$ is the pointwise product, ${\mathcal{P}}\in \mathbb{C}^{N_1 \times N_2 \times \cdots \times N_d}$ is the $d$-way observation tensor{\footnote{This model subsumes a number of signal processing systems. For example in harmonic retrieval~\cite{chi2015compressive}, ${\mathcal{X}}$ is the data tensor and ${\mathcal{P}}$ is a sparse sampling tensor, which observes data tensor uniformly at random. Furthermore, in communication and passive radar systems, ${\mathcal{X}}$ is the channel matrix and ${\mathcal{P}}$ is the data symbol matrix~\cite{li2019multi,zheng2018adaptive}. }}, and ${\mathcal{N}} \in \mathbb{C}^{N_1 \times N_2 \times \cdots \times N_d} $ is a white, complex circularly symmetric Gaussian noise tensor.

In this paper, the following prior knowledge is assumed on the unknown frequencies 
\begin{align}
\label{eq:priori}
[f_{i,1},...,f_{i,r}]^T \in {\mathbb{F}}_{i}^{r \times 1} = [f_{L,i},f_{H,i}]^{r \times 1} \in {[0,1)}^{r \times 1},~i=1,...,d,
\end{align}
where $[f_L,f_H]$ denotes a closed interval as usual if $f_L< f_H$, otherwise we define $[f_L,f_H]\triangleq {[0,1)} \backslash (f_H,f_L)$. Then we aim to estimate $\bm f_{\ell} \triangleq [f_{1,\ell},...,f_{d,\ell}]^T \in {[0,1)}^{d\times 1},~\ell = 1,...,r$ from the measurements ${\mathcal{Y}}$ in \eqref{eq:y-1}.

\section{Proposed Methods Based on MD-FS Vandermonde Decomposition}

In this section, we present the proposed MD spectral super-resolution methods based on the MD-FS Vandermonde decomposition. First the MD-FS atomic norm is introduced to formulate the MD spectral super-resolution problems. Then by using the MD-FS Vandermonde decomposition result on ML block Toeplitz matrices, we convert the MD-FS atomic norm optimization problems into SDP formulations.

\subsection{MD Spectral Super-resolution Based on MD-FS Atomic Norm}

Inspired by the FS atomic norm approach~\cite{yang2018frequency,mishra2015spectral}, we define the MD-FS atomic set as the collection of all MD complex sinusoids:
\begin{align}
{\mathbb{A}}({\mathbb{F}})\triangleq\{  {\cal A}(\bm f): f_i \in {\mathbb{F}}_{i},~i=1,...,d \},
\end{align}
where ${\cal A}(\bm f)$ is given by \eqref{MDsimusoid}.
Then, the MD-FS atomic norm with respect to ${\cal X}$ in \eqref{eq:tensorX2} is defined as follows.
\begin{definition}
	The MD-FS atomic norm for ${\cal X}$ in \eqref{eq:tensorX2} is 
	\begin{align}
\label{eq:2DAN-definition}
\|{\cal X}\|_{{\mathbb{A}}({\mathbb{F}})}\triangleq &~ \inf \{ \chi>0: {\cal X} \in \chi {\rm conv}({\mathbb{A}}({\mathbb{F}})) \} \nonumber \\
=&~\inf_{\substack{f_{i,\ell} \in {\mathbb{F}}_{i}, i=1,...,d \\ \sigma_{\ell} \in \mathbb{C}}} \left\{     \sum_{\ell} |\sigma_{\ell}| : {\cal X} =  \sum_{\ell} \sigma_{\ell}  {\cal A}(\bm f_{\ell})  \right\}.
\end{align}
\end{definition}

On this basis, in the absence of noise, i.e. $\cal{N} = \emptyset$ in \eqref{eq:y-1}, then our MD frequency estimation problem can be formulated as the following convex constrained form~\cite{chen2001atomic,donoho2006compressed}
\begin{align}
\label{eq:AN-problem-1}
\widehat{\cal X} = \arg \min_{{\cal X} } \| {\cal X}\|_{{\mathbb{A}}({\mathbb{F}})},~{\text{s.t.}}~  {\mathcal{Y}} = {\mathcal{P}} \odot {\mathcal{X}}.
\end{align}
Moreover, with noise $\cal{N}$ in \eqref{eq:y-1}, the problem can be formulated as the following convex unconstrained form~\cite{chen2001atomic}
\begin{align}
\label{eq:AN-problem-2}
\widehat{\cal X} = \arg \min_{{\cal X} } \frac{1}{2} \| {\mathcal{Y}} - {\mathcal{P}} \odot {\cal X} \|_F^2 + \lambda \| {\cal X}\|_{{\mathbb{A}}({\mathbb{F}})},
\end{align}
where $\lambda > 0$ is a weight factor. Once ${\cal X}$ is obtained from \eqref{eq:AN-problem-1} or \eqref{eq:AN-problem-2}, one way to determine the frequencies $\bm f_{1},..., \bm f_{r}$ and complex gains $\bm \sigma \triangleq [\sigma_{1},...,\sigma_{r}]^T$ is to use the MD MUltiple SIgnal Classifier (MD-MUSIC)~\cite{swindlehurst1992performance} algorithm with ${\cal X}$ as an input. In particular, the MD-MUSIC method determines the frequencies by locating the poles in the spectrum and estimates the complex gains by the least-squares method with the estimated frequencies. Alternatively, one can obtain the frequencies from the dual solutions of the problems. Define the dual norm of $\|\cdot\|_{{\mathbb{A}}({\mathbb{F}})}$ as
\begin{align}
\label{eq:dual-norm}
\| {\cal V} \|_{{\mathbb{A}}({\mathbb{F}})}^* \triangleq \sup_{\|{\cal X}\|_{{\mathbb{A}}({\mathbb{F}})} \leq 1 } \langle  {\cal V},  {\cal P} \odot {\cal X} \rangle_{\Re},
\end{align} 
where $\langle {\cal V}, {\cal X} \rangle_{\Re} = \Re(\langle {\cal V}, {\cal X} \rangle)$.
Following the standard Lagrangian analysis~\cite{chandrasekaran2012convex}, the dual problems of \eqref{eq:AN-problem-1} and \eqref{eq:AN-problem-2} are respectively given by
\begin{align}
\label{eq:dual-problem-1}
&\max_{ {\cal V} } \langle {\cal V}, {\cal Y} \rangle_{\Re}, ~{\text{s.t.}}~\| {\cal V} \|_{{\mathbb{A}}({\mathbb{F}})}^* \leq 1, \\
\label{eq:dual-problem-2}
&\max_{ {\cal V} } \langle {\cal V}, {\cal Y} \rangle_{\Re} - \frac{1}{2} \| {\cal V} \|_F^2, ~{\text{s.t.}}~\| {\cal V} \|_{{\mathbb{A}}({\mathbb{F}})}^* \leq \lambda.
\end{align}
Solving the dual problems is equivalent to solving the primal problems, and most solvers can directly return dual optimal solutions when solving the primal problems. We can then obtain the frequencies from the dual solutions according to the following lemma since the strong duality holds, which is an extension of the 1D results given by the Proposition II.4 in~\cite{tang2013compressed}.
\begin{lemma} 
Suppose ${\widehat{\cal X}} = \sum_{\ell = 1}^{\widehat r} \widehat\sigma_{\ell}  {\cal A}(\bm{\widehat{f}}_{\ell})$ is the primal solution, then the dual polynomials $Q(\bm f_{\ell}) \triangleq \langle {\widehat{\cal V}}, {\cal P} \odot {\cal A}(\bm f_{\ell}) \rangle$ of \eqref{eq:dual-problem-1} and \eqref{eq:dual-problem-2} respectively satisfy
\begin{align}
\label{eq:dual-solution-1}
Q(\bm{\widehat{f}}_{\ell}) =&~ \frac{\widehat \sigma_{\ell}}{|\widehat \sigma_{\ell}|},~\ell = 1,...,\widehat r, \\
\label{eq:dual-solution-2}
Q(\bm{\widehat{f}}_{\ell}) =&~ \lambda \frac{\widehat \sigma_{\ell}}{|\widehat \sigma_{\ell}|},~\ell = 1,...,\widehat r.
\end{align}
\end{lemma}
Hence, the frequencies in \eqref{eq:AN-problem-1} and \eqref{eq:AN-problem-2} can be determined by identifying points where the dual polynomials have moduli $1$ and $\lambda$, respectively. Then, the complex gains can be estimated by the least-squares method with the estimated frequencies. Since the MD-FS atomic norm in \eqref{eq:AN-problem-1} and \eqref{eq:AN-problem-2} is essentially a semi-infinite program, it cannot be directly solved. We show in the following subsections how to solve \eqref{eq:AN-problem-1} and \eqref{eq:AN-problem-2} based on the MD-FS Vandermonde decomposition.



\subsection{MD-FS Vandermonde Decomposition of ML Block Toeplitz Matrices}


Recall that $\bm a_i^{\ell} \triangleq \bm s(f_{i,\ell},N_{i}),~i = 1,...,d, ~\ell = 1,...,r$, define 
\begin{align}
\label{eq:veca}
\bm a(\bm f_{\ell}) \triangleq {\rm vec}(\bm a_1^{\ell} \otimes \bm a_2^{\ell} \otimes \cdots  \otimes \bm a_d^{\ell}) = {\rm vec}({\cal A}(\bm f_{\ell})).
\end{align}
 Hence, for any $\ell$, we have that $\bm a(\bm f_{\ell}) \bm a^H(\bm f_{\ell})$ has the form in \eqref{eq:Td-vecS}, which is a rank-1 $d$-level block Toeplitz matrix. To solve an MD-FS atomic norm minimization problem, the basic idea is to find an ML block Toeplitz matrix, and convert the problem of minimizing $\sum_{\ell} |\sigma_{\ell}|$ in the atomic norm (the convex relaxation of minimizing the number of $d$-dimensional sinusoids) to minimizing the trace of the ML block Toeplitz matrix (the convex relaxation of minimizing the rank of the block Toeplitz matrix). We hence present the MD-FS Vandermonde decomposition result of ML block Toeplitz matrices, and then we can convert the atomic norm in \eqref{eq:AN-problem-1} and \eqref{eq:AN-problem-2} into a semidefinite program based on the Vandermonde decomposition result of the Toeplitz matrix.

%

\begin{theorem} \label{the: MDVD} (Theorem 1, \cite{yang2016vandermonde2}) Assume that $\bm T^d$ is a PSD $d$-level block Toeplitz matrix with $d \geq 1$ and $r = {\rm rank}(\bm T^d) < \min_i N_i$. Then, $\bm T^d$ can be decomposed as
\begin{align}
\label{eq:VSV}
\bm{T}^d =&~ \bm A \bm \Sigma \bm A^H = \sum_{\ell=1}^{r} \sigma_{\ell} \bm{a}\left(\bm{f}_{\ell}\right) \bm{a}^{H}\left(\bm{f}_{\ell}\right), \\
\label{eq:A}
\text{with}~\bm{A} =&~ [\bm a(\bm{f}_{1}),...,\bm a(\bm{f}_{r})] \in \mathbb{C}^{N_D\times r}, \\
\label{eq:Sigma}
\bm \Sigma =&~ {\rm diag}(\bm \sigma) = {\rm diag}([\sigma_{1},...,\sigma_{r}]^T) \in \mathbb{C}^{r \times r},
\end{align}
where $\sigma_{\ell}>0,~\bm f_{\ell},~\ell = 1,...,r$ are distinct points points in ${[0,1)}^{d\times 1}$, and the $(d + 1)$-tuples $(\bm f_{\ell},\sigma_{\ell}),~\ell = 1,...,r$ are unique.
\end{theorem}

    The above theorem shows that once $r = {\rm rank}(\bm T^d) < \min_i N_i$ holds, the $d$-level block Toeplitz matrix ${\bm T}^d$ has the MD Vandermonde decomposition in \eqref{eq:VSV} if ${\bm T}^d \succeq 0$. In order to combine the interval information into MD Vandermonde decomposition. We give a property of trigonometric polynomials in the following lemma. The proof is given in Appendix A.

\begin{lemma} \label{lem: ptp} For the trigonometric polynomials in \eqref{eq:htps}, if
\begin{align}
\label{eq:r0}
r_{0,i} =&~ -2 \cos[\pi(f_{H,i}-f_{L,i})] {\rm sign}(f_{H,i}-f_{L,i}), \\
\label{eq:r1}
r_{1,i} =&~ e^{i\pi(f_{L,i}+f_{H,i})} {\rm sign}(f_{H,i}-f_{L,i}).
\end{align}
then, when $f_{L,i} \neq f_{H,i}$, 
\begin{align}
\label{eq:g1}
g_i(f_i) =&~ r_{1,i} x_i^{-1} + r_{0,i} + r_{-1,i} x_i \nonumber \\
=&~  r_{0,i} + 2\Re\{r_{1,i}e^{-i2\pi f_i}\},~i=1,...,d
\end{align}
are always positive on $(f_{L,i},f_{H,i})$ and negative on $(f_{H,i},f_{L,i})$.
\end{lemma}

The above lemma shows that we can restrict the frequencies in given intervals by setting $g_i(f_i)\geq 0$. To this end, we introduce the MD-FS Vandermonde decomposition of $d$-level block Toeplitz matrices in the following proposition.

\begin{proposition} \label{thm: FSVD} For a $d$-level block Toeplitz matrix $\bm T^d \in \mathbb{C}^{N_D\times N_D}$ with $d\geq 2$, if $r = {\rm rank}(\bm T^d) < \min_i N_i$, then given ${\mathbb{F}}_{i} \in {[0,1)},~i=1,...,d$, it has an MD-FS Vandermonde decomposition as in \eqref{eq:VSV} with $f_{i,\ell} \in {\mathbb{F}}_{i},~\ell = 1,...,r,~i=1,...,d$, if and only if 
\begin{align}
\label{eq:T}
\bm T^d &\succeq \bm 0, \\
\label{eq:Tp}
\bm T_{g_i}^d &\succeq \bm 0,~i=1,...,d,
\end{align}
where $g_i$ and $\bm T_{g_i}^d$ are defined by \eqref{eq:r0}-\eqref{eq:g1} and \eqref{dTgd}, respectively.
\end{proposition}

\begin{IEEEproof}
We first prove the sufficient condition. By \eqref{eq:T} and Theorem~\ref{the: MDVD}, $\bm T^d$ has an MD Vandermonde decomposition as in \eqref{eq:VSV}. Hence, we need to prove $f_{i,\ell} \in {\mathbb{F}}_{i},~\ell = 1,...,r,~i=1,...,d$ under the additional conditions \eqref{eq:Tp}. For the MD Vandermonde decomposition in \eqref{eq:VSV}, we have
\begin{align}
&{\cal B}^d(m_1-n_1,...,m_d-n_d) = \bm T^d(m_1,n_1;...;m_d,n_d) \nonumber \\
&= \sum_{\ell=1}^r \sigma_{\ell} e^{i2\pi(m_1-n_1)f_{1,\ell}} \cdots e^{i2\pi(m_d-n_d)f_{d,\ell}},
\end{align}
which shows that for $i = 1,...,d$
\begin{align}
\bm T_{g_i}^d(m_1,n_1;...;m_d,n_d) =&~ \sum_{k=-1}^{1} r_{k,i} {\cal B}^d(m_1-n_1,...,m_i-n_i-k,...,m_d-n_d)  \nonumber \\
=&~ \sum_{k=-1}^{1} r_{k,i} \sum_{\ell=1}^r \sigma_{\ell} e^{i2\pi(m_1-n_1)f_{1,\ell}} \cdots e^{i2\pi(m_i-n_i-k)f_{i,\ell}} \cdots e^{i2\pi(m_d-n_d)f_{d,\ell}} \nonumber \\
=&~ \sum_{\ell=1}^r \sigma_{\ell} e^{i2\pi(m_1-n_1)f_{1,\ell}} \cdots e^{i2\pi(m_d-n_d)f_{d,\ell}}  \sum_{k=-1}^{1} r_{k,i} e^{-i2\pi k f_{i,\ell}} \nonumber \\
=&~ \sum_{\ell=1}^r \sigma_{\ell} g_i(f_{i,\ell}) e^{i2\pi(m_1-n_1)f_{1,\ell}} \cdots e^{i2\pi(m_d-n_d)f_{d,\ell}},\\
&~m_1,n_1 = 1,...,N_1-1;...;m_d,n_d=1,...,N_d-1, \nonumber
\end{align}
and hence,
\begin{align}
\label{eq:AdA}
\bm T_{g_i}^d &= \bm{\bar{A}} {\rm diag}\left([\sigma_1 g_i(f_{i,1}),...,\sigma_{r} g_i(f_{i,r})]^T \right) \bm{\bar{A}}^H, \\
\text{with}~\bm{\bar{A}} &\triangleq [\bm{\bar{a}}(\bm f_{1}),...,\bm{\bar{a}}(\bm f_{r})] \in \mathbb{C}^{{\bar N}_{D}\times r}, \\
\bm{\bar{a}}(\bm f_{\ell}) &\triangleq {\rm vec}( \bm{\bar{a}}_1^{\ell}  \otimes \cdots \otimes \bm{\bar{a}}_d^{\ell} ) \in \mathbb{C}^{{\bar N}_{D}\times 1},~\ell = 1,...,r, \\
\bm{\bar{a}}_i^{\ell} & \triangleq \bm s(f_{i,\ell},N_{i}-1) \in \mathbb{C}^{(N_i-1)\times 1},~i = 1,...,d.
\end{align}
Since $r < \min_i N_i$, $\bm{\bar{A}}$ has full column rank. Then, by noting \eqref{eq:AdA} and \eqref{eq:Tp} we have for $i=1,...,d$
\begin{align}
\label{eq:ATpA}
{\rm diag}\left([\sigma_1 g_i(f_{i,1}),...,\sigma_{r} g_i(f_{i,r})]^T \right) = \bm{\bar{A}}^{\dagger} \bm T_{g_i}^d \bm{\bar{A}}^{\dagger H} \geq 0,
\end{align}
where $(\cdot)^{\dagger}$ denotes the matrix pseudo-inverse operator. \eqref{eq:ATpA} implies that $\sigma_{\ell} g_i(f_{i,\ell})\geq 0,~i=1,...,d$, it immediately follows that $g_i(f_{i,\ell})\geq 0,~i=1,...,d$ since $\sigma_{\ell}>0,~\ell = 1,...,r$. By noting Lemma~\ref{lem: ptp} we finally have $f_{i,\ell} \in {\mathbb{F}}_{i},~\ell = 1,...,r$.

Next we prove the necessary condition, which is trivial. Given $\bm{T}^d$ in \eqref{eq:VSV} with $f_{i,\ell} \in {\mathbb{F}}_{i},~\ell = 1,...,r,~i=1,...,d$. We have \eqref{eq:T} holds since $\sigma_{\ell}>0,~\ell = 1,...,r$. Moreover, \eqref{eq:Tp} also holds since we have $g_i(f_{i,\ell})\geq 0,~i=1,...,d,~\ell = 1,...,r$ in \eqref{eq:ATpA} by noting Lemma~\ref{lem: ptp}. Therefore we complete the proof.
\end{IEEEproof}

It is noteworthy that the MD-FS Vandermonde decomposition result can be extended to the multiple frequency band case, as stated by the following corollary. 

\begin{corollary} \label{cor: FSVD} For a $d$-level block Toeplitz matrix $\bm T^d({\cal B}^d) \in \mathbb{C}^{N_D\times N_D}$, if $r = {\rm rank}(\bm T^d({\cal B}^d)) < \min_i N_i$, it admits an MD-FS Vandermonde decomposition as in \eqref{eq:VSV} with $f_{i,\ell} \in \bigcup_{j}^J {\mathbb{F}}_{i,j},~\ell = 1,...,r,~i=1,...,d$ where ${\mathbb{F}}_{i,j} = [f_{L,i,j},f_{H,i,j}] \in {[0,1)},~j=1,...,J,~i=1,...,d$, if and only if there exist $d$-way tensors ${\cal B}^d_j,~j = 1,...,J$ satisfying
\begin{align}
\label{eq:cor-B}
&~\sum_{j=1}^J {\cal B}^d_j = {\cal B}^d, \\
\label{eq:cor-rank}
&~\sum_{j=1}^J {\rm rank}(\bm T^d({\cal B}^d_j)) = r, \\
\label{eq:cor-T-c}
&~\bm T^d({\cal B}^d_j) \succeq  \bm 0, \\
\label{eq:cor-Tp-c}
&~\bm T_{g_{i,j}}^d({\cal B}^d_j) \succeq  \bm 0,~i = 1,...,d,
\end{align}
where $g_{i,j}$ is defined with respect to $[f_{L,i,j},f_{H,i,j}]$.
\end{corollary}
\begin{IEEEproof}
We first prove the sufficient condition. Suppose $r_j = {\rm rank}(\bm T^d({\cal B}^d_j)),~j=1,...,J$. For each $j$, we have $r_j < \min_i N_i$, if \eqref{eq:cor-T-c} and \eqref{eq:cor-Tp-c} hold, then there exists an MD Vandermonde decomposition of $\bm T^d({\cal B}^d_j)$ with $f_{i,\ell} \in {\mathbb{F}}_{i,j},~\ell = 1,...,r,~i=1,...,d$ according to Proposition~\ref{thm: FSVD} as
\begin{align}
\label{eq:VSVj}
\bm T^d({\cal B}^d_j) = \bm A_j {\rm diag}(\bm \sigma_j) \bm A_j^H,
\end{align}
where $\bm A_j \in \mathbb{C}^{N_D \times r_j}$ and $\bm \sigma_j \in \mathbb{C}^{r_j \times 1}$. If we set $\bm A$ and $\bm \Sigma$ in \eqref{eq:A} and \eqref{eq:Sigma} as
\begin{align}
\bm A =&~ [\bm A_1,...,\bm A_J] \in \mathbb{C}^{N_D \times r}, \\
\bm \Sigma =&~ {\rm diag}([\bm \sigma_1^T,...,\bm \sigma_J^T]^T) \in \mathbb{C}^{r \times r},
\end{align}
then we have $\bm T^d({\cal B}^d) = \bm A \bm \Sigma \bm A^H$ in \eqref{eq:VSV} with $f_{i,\ell} \in \bigcup_{j}^J {\mathbb{F}}_{i,j},~\ell = 1,...,r,~i=1,...,d$ since $\sum_{j=1}^J {\cal B}^d_j = {\cal B}^d$ and $\sum_{j=1}^J r_j = r$ hold.

Next we prove the necessary condition. For any $\bm T^d({\cal B}^d) = \bm A \bm \Sigma \bm A^H$ in \eqref{eq:VSV} with $r = {\rm rank}(\bm T^d({\cal B}^d)) < \min_i N_i$ and $f_{i,\ell} \in \bigcup_{j}^J {\mathbb{F}}_{i,j},~\ell = 1,...,r,~i=1,...,d$, we can decompose the frequencies $\bm f_{\ell},~\ell = 1,...,r$ into $J$ groups $\omega^{(1)},...,\omega^{(J)}$ with respect to ${\mathbb{F}}_{i,1},...,{\mathbb{F}}_{i,J}$, such that for each ${\ell}$, $\bm f_{\ell} \in \omega^{(j)}$ for some $j$, with $f_{i,\ell} \in {\mathbb{F}}_{i,j},~i=1,...,d$. 
Accordingly, ${\cal B}^d$ can be decomposed into $J$ tensors ${\cal B}^d_j,~j=1,...,J$ with respect to $\omega^{(1)},...,\omega^{(J)}$. Then \eqref{eq:cor-B} and \eqref{eq:cor-rank} hold. And for each $\bm T^d({\cal B}^d_j)$, we further have \eqref{eq:cor-T-c} and \eqref{eq:cor-Tp-c} hold according to Proposition~\ref{thm: FSVD}. Then we complete the proof.
\end{IEEEproof}

\subsection{SDP Formulation of MD-FS Atomic Norm}

Denote $\bm y \triangleq {\rm vec}({\mathcal{Y}}) \in \mathbb{C}^{N_D\times 1}$, $\bm x \triangleq {\rm vec}({\mathcal{X}}) \in \mathbb{C}^{N_D\times 1}$ and $\bm \Phi \triangleq {\rm diag}({\rm vec}({\mathcal{P}})) \in \mathbb{C}^{N_D\times N_D}$. Under the condition ${\rm rank}(\bm T^d) < \min_i N_i $, the MD-FS atomic norm minimizations in \eqref{eq:AN-problem-1} and \eqref{eq:AN-problem-2} can be converted to SDP formulations by applying the MD-FS Vandermonde decomposition as stated by the following proposition.

\begin{proposition} \label{prop: FSAN} For the MD-FS atomic norm defined in~\eqref{eq:2DAN-definition}, we have that
\begin{align}
\label{eq:FSAN-thm}
\|{\cal X}\|_{{\mathbb{A}}({\mathbb{F}})} \geq &~ \min_{{\cal B}^d, {t}}  \frac{1}{2N_D} {\rm Tr}(\bm T^d({\cal B}^d)) + \frac{1}{2} {t}, \\
\label{eq:FSAN-constraint}
&~{\text{s.t.}}~\left[ {\begin{array}{*{20}{c}}
	\bm T^d({\cal B}^d) & \bm x \\
	\bm x^H & {t}
	\end{array}} \right] \succeq 0,~\bm T_{g_i}^d({\cal B}^d) \succeq 0,~i=1,...,d,
\end{align}
where $g_i$ and $\bm T_{g_i}^d$ are defined by \eqref{eq:r0}-\eqref{eq:g1} and \eqref{dTgd}, respectively. And if ${\rm rank}(\bm T^d({\cal B}^d)) < \min_i N_i$, we further have $\|{\cal X}\|_{{\mathbb{A}}({\mathbb{F}})}$ equals to the right-hand side of \eqref{eq:FSAN-thm}.
\end{proposition}

\begin{IEEEproof} Denote the value of the right-hand side of \eqref{eq:FSAN-thm} by ${\rm SDP}{(\bm x)}$. Let ${\cal X} = \sum_{\ell} \sigma_{\ell} {\cal A}(\bm f_{\ell})$, where $\sigma_{\ell} = |\sigma_{\ell}| e^{ i \theta_{\ell}}$, be an MD-FS atomic decomposition on ${\mathbb{F}}_{i},~i=1,...,d$. We first show that the constraints in \eqref{eq:FSAN-constraint} hold. We have $\bm T_{g_i}^d \succeq 0,~i=1,...,d$ since by \eqref{eq:AdA}
\begin{align}
\label{eq:AdA2}
\bm T_{g_i}^d = \bm{\bar{A}} {\rm diag}\left([\sigma_1 g_i(f_{i,1}),...,\sigma_{r} g_i(f_{i,r})]^T \right) \bm{\bar{A}}^H,
\end{align}
with $\sigma_{\ell}>0,~\ell = 1,...,r$ and $g_i(f_{i,\ell})\geq 0,~i=1,...,d,~\ell = 1,...,r$ by noting Lemma~\ref{lem: ptp}. Moreover, since $\bm x = {\rm vec}({\cal X}) = \sum_{\ell} \sigma_{\ell} {\rm vec}({\cal A}(\bm f_{\ell})) = \sum_{\ell} \sigma_{\ell} \bm a(\bm f_{\ell})$ and $\bm T^d = \sum_{\ell} |\sigma_{\ell}|{\bm a(\bm f_{\ell})}{\bm a(\bm f_{\ell})}^H$ by \eqref{eq:Td-vecS}, then
\begin{eqnarray}
\left[ {\begin{array}{*{20}{c}}
	\bm T^d & \bm x\\
	\bm x^H& {\sum\limits_{\ell} |\sigma_{\ell}|}
	\end{array}} \right]
	=
	\sum_{\ell} |\sigma_{\ell}| \left[ {\begin{array}{*{20}{c}}
	{\bm a(\bm f_{\ell})}\\
	{e^{i\theta_{\ell}}}
	\end{array}} \right] 
	\left[ {\begin{array}{*{20}{c}}
	{\bm a(\bm f_{\ell})}\\
	{e^{i\theta_{\ell}}}
	\end{array}} \right]^H  \succeq 0.
\end{eqnarray}
Now since ${\rm SDP}{(\bm x)}$ is the solution to the minimization problem \eqref{eq:FSAN-thm}-\eqref{eq:FSAN-constraint}, we have
\begin{eqnarray}
\label{eq:SDP-inequality-1}
{\rm SDP}{(\bm x)}\leq \frac{1}{2N_D}{\rm{Tr}}(\bm T^d) + \frac{1}{2}  {\sum_{\ell} |\sigma_{\ell}|} = {\sum_{\ell} |\sigma_{\ell}|}.
\end{eqnarray}
Since ${\rm SDP}{(\bm x)}\leq {\sum_{\ell} |\sigma_{\ell}|}$ holds for any decomposition of ${\cal X}$, we conclude that ${\rm SDP}{(\bm x)}\leq \|{\cal X}\|_{{\mathbb{A}}({\mathbb{F}})}$.

Now denote the optimal solution to \eqref{eq:FSAN-thm}-\eqref{eq:FSAN-constraint} as ${\widehat{\cal B}}^d$ and $\widehat {t}$, then we have $\bm T^d({\widehat{\cal B}}^d) \succeq 0$ and 
\begin{eqnarray}
\label{eq:TtxxH}
\bm T^d({\widehat{\cal B}}^d) \succeq {\widehat{t}}^{-1}\bm x \bm x^H
\end{eqnarray}
by the Schur complement condition. Note that $\bm\Theta = \left[ {\begin{array}{*{20}{c}}
	{\bm{T}}^d & {\bm{x}}\\
	{\bm x}^H& {t}
	\end{array}} \right] \succeq 0$ 
implies that ${\bm{x}} \in {\rm span}({\bm{T}}^d)$. To see this, suppose otherwise ${\bm{x}} = {\bm{x}}^{\parallel} + {\bm{x}}^{\perp}$ where ${\bm{x}}^{\parallel} \in {\rm span}({\bm{T}}^d)$ and ${\bm{x}}^{\perp} \in {\rm null}({\bm{T}}^d)$. Let ${\bm{q}} \in {\rm null}({\bm{T}}^d)$, then we have $[{\bm{q}}^H,  0] \bm \Theta \left[ {\begin{array}{*{20}{c}}
	{\bm{q}} \\
	{0}
	\end{array}} \right] = 0$. Since $\bm\Theta \succeq 0$, we can write $\bm\Theta = \bm R \bm R^H$. Hence
\begin{align}	
[{\bm{q}}^H,  0] \bm R = \bm 0 \Rightarrow [{\bm{q}}^H,  0] \bm R \bm R^H = \bm 0 = [{\bm{q}}^H,  0] \bm \Theta \Rightarrow {\bm{q}}^H {\bm{x}} = 0 = {\bm{q}}^H ({\bm{x}}^{\parallel} + {\bm{x}}^{\perp}) = {\bm{q}}^H {\bm{x}}^{\perp}.
\end{align}
Therefore, ${\bm{x}}^{\perp} = \bm 0$, i.e., ${\bm{x}} \in {\rm span}({\bm{T}}^d)$.

If ${\rm rank}( \bm T^d({\widehat{\cal B}}^d) ) < \min_i N_i$, we have $\bm T^d({\widehat{\cal B}}^d) = \bm{\widehat A} \bm{\widehat \Sigma} \bm{\widehat A}^H$ according to Proposition~\ref{thm: FSVD}. Then we can write
\begin{align}
\label{eq:xAw}
\bm x = \sum_{\ell} w_{\ell} \bm a(\bm{\widehat{f}}_{\ell})= \bm{\widehat A} \bm w,
\end{align}
for some complex coefficient vector $\bm w$. Hence, we have $\bm{\widehat A} \bm{\widehat \Sigma} \bm{\widehat A}^H \succeq {\widehat{t}}^{-1} \bm{\widehat A} \bm w \bm w^H \bm{\widehat A}^H$ from \eqref{eq:TtxxH}. Since $\bm{\widehat A}$ has full column-rank, let $\bm z$ be such that $(\bm{\widehat A}^H \bm z)_j = \frac{w_j}{|w_j|}$, we have
\begin{eqnarray}
{\rm Tr}(\bm{\widehat \Sigma})  = \bm z^H \bm{\widehat A} \bm{\widehat \Sigma} \bm{\widehat A}^H \bm z \geq {\widehat{t}}^{-1}\bm z^H \bm{\widehat A} \bm w \bm w^H \bm{\widehat A}^H \bm z = {\widehat{t}}^{-1} \left (\sum_{\ell}|w_{\ell}| \right )^2.
\end{eqnarray}
It therefore follows that
\begin{align}
\frac{1}{2N_D}{\rm{Tr}}(\bm T^d({\widehat{\cal B}}^d)) + \frac{1}{2} \widehat {t} =\frac{1}{2} {\rm{Tr}}(\bm{\widehat \Sigma}) + \frac{1}{2} \widehat {t}  \geq \sqrt{{\rm{Tr}}(\bm{\widehat \Sigma}) \widehat {t} } \geq  \sum_{\ell}|w_{\ell}| \geq \|{\cal X}\|_{{\mathbb{A}}({\mathbb{F}})},
\end{align}
which is equivalent to ${\rm SDP}{(\bm x)} \geq \|{\cal X}\|_{{\mathbb{A}}({\mathbb{F}})}$. This together with \eqref{eq:SDP-inequality-1} leads to $\|{\cal X}\|_{{\mathbb{A}}({\mathbb{F}})} = {\rm SDP}{(\bm x)}$ if ${\rm rank}(\bm T^d) < \min_i N_i$, which completes the proof.
\end{IEEEproof}

Note that Proposition~\ref{prop: FSAN} can also be extended to the multiple frequency band case as in the following corollary by applying Corollary~\ref{cor: FSVD}, the proof of which is straightforward by following the proof of Proposition~\ref{prop: FSAN} and thus is omitted.
\begin{corollary} 
\label{coro: FSAN} For the multiple frequency band MD-FS atomic norm defined as
\begin{align}
\label{eq:multiple-band-FSAN-definition}
\|{\cal X}\|_{{\mathbb{A}}({\mathbb{F}}_M)}\triangleq  \inf_{\substack{f_{i,\ell} \in \bigcup_{j}^J {\mathbb{F}}_{i,j}, i=1,...,d \\ \sigma_{\ell} \in \mathbb{C}}} \left\{     \sum_{\ell} |\sigma_{\ell}| : {\cal X} =  \sum_{\ell} \sigma_{\ell}  {\cal A}(\bm f_{\ell})  \right\},
\end{align}
we have
\begin{align}
\label{eq:FSAN-cor}
\|{\cal X}\|_{{\mathbb{A}}({\mathbb{F}}_M)} \geq &~ \min_{{\cal B}^d, {t}}  \frac{1}{2N_D} \sum_{j=1}^J {\rm Tr}(\bm T^d({\cal B}^d_j)) + \frac{1}{2} {t}, \\
&~{\text{s.t.}}~\left[ {\begin{array}{*{20}{c}}
	\bm T^d(\sum_{j=1}^J {\cal B}^d_j) & \bm x \\
	\bm x^H & {t}
	\end{array}} \right] \succeq 0, \nonumber \\
&~~~~~~\bm T_{g_{i,j}}^d({\cal B}^d_j) \succeq  \bm 0,~i = 1,...,d,~j=1,...,J, \nonumber
\end{align}
where $g_{i,j}$ is defined with respect to ${\mathbb{F}}_{i,j} = [f_{L,i,j},f_{H,i,j}]$. Moreover, if $~\sum_{j=1}^J {\rm rank}(\bm T^d({\cal B}^d_j)) < \min_i N_i$, we further have $\|{\cal X}\|_{{\mathbb{A}}({\mathbb{F}}_M)}$ equals to the right-hand side of \eqref{eq:FSAN-cor}.
\end{corollary} 

By applying Proposition~\ref{prop: FSAN} we can approximately{\footnote{Although the SDPs are approximations, simulation results show that the performance is good even if the condition ${\rm rank}(\bm T^d) < \min_i N_i$ is not satisfied.} convert \eqref{eq:AN-problem-1} and \eqref{eq:AN-problem-2} into the following SDPs, which are exact under the condition ${\rm rank}(\bm T^d({\cal B}^d)) < \min_i N_i$:
\begin{align}
\label{eq:SDP-problem-1}
& \min_{\bm x, {\cal B}^d, {t} } \frac{1}{2N_D} {\rm Tr}(\bm T^d({\cal B}^d)) + \frac{1}{2} {t}, \\
&~{\text{s.t.}}~\bm y = \bm \Phi \bm x,~\left[ {\begin{array}{*{20}{c}}
	\bm T^d({\cal B}^d) & \bm x \\
	\bm x^H & {t}
	\end{array}} \right] \succeq 0,~\bm T_{g_i}^d({\cal B}^d) \succeq 0,~i=1,...,d, \nonumber \\
\label{eq:SDP-problem-2}
& \min_{\bm x, {\cal B}^d, {t} } \frac{1}{2} \| \bm y - \bm \Phi \bm x \|_2^2 + \frac{\lambda}{2N_D} {\rm Tr}(\bm T^d({\cal B}^d)) + \frac{\lambda}{2} {t}, \\
&~{\text{s.t.}}~\left[ {\begin{array}{*{20}{c}}
	\bm T^d({\cal B}^d) & \bm x \\
	\bm x^H & {t}
	\end{array}} \right] \succeq 0,~\bm T_{g_i}^d({\cal B}^d) \succeq 0,~i=1,...,d. \nonumber
\end{align}
After $\bm x$ is obtained from \eqref{eq:SDP-problem-1} or \eqref{eq:SDP-problem-2}, as mentioned in Section III-A, the frequencies can be determined by the MD-MUSIC algorithm or solving the dual problems of \eqref{eq:SDP-problem-1} and \eqref{eq:SDP-problem-2}. Define $\| \bm \nu \|_{{\cal{A}}({\mathbb{F}})} \triangleq \| {\cal V} \|_{{\cal{A}}({\mathbb{F}})}$ with $\bm \nu  = {\rm vec}({\cal V})$. Following the analysis in~\cite{tang2013compressed,dumitrescu2007positive}, we can write the dual problems of \eqref{eq:SDP-problem-1} and \eqref{eq:SDP-problem-2} based on \eqref{eq:dual-problem-1} and \eqref{eq:dual-problem-2} respectively as
\begin{align}
\label{eq:SDP-dual-1}
&\max_{\bm \nu, \bm Q, \bm Q_{g_i}, i=1,...,d} \langle \bm \nu, \bm y \rangle_{\Re}, \\
&{\text{s.t.}}~ \langle \bm Q, \bm \Upsilon_{\bm p} \rangle + \sum_{i=1}^{d} \langle \bm Q_{g_i}, \bm \Upsilon_{g_i,\bm p} \rangle  = \delta_{\bm p}, ~\bm p = (p_1,...,p_d),~p_i  \in \{-N_i+1,..., N_i-1\}, \nonumber \\
&~~~\left[ {\begin{array}{*{20}{c}}
	\bm Q & \bm \Phi \bm \nu \\
	(\bm \Phi \bm \nu)^H & {1}
	\end{array}} \right] \succeq 0,~\bm Q_{g_i} \succeq 0,~i = 1,...,d, \nonumber
\end{align}
\begin{align}
\label{eq:SDP-dual-2}
\text{and}~&\max_{\bm \nu, \bm Q, \bm Q_{g_i}, i=1,...,d}  \langle \bm \nu, \bm y \rangle_{\Re} - \frac{1}{2} \| \bm \nu \|_2^2, \\
&{\text{s.t.}}~ \langle \bm Q, \bm \Upsilon_{\bm p} \rangle + \sum_{i=1}^{d} \langle \bm Q_{g_i}, \bm \Upsilon_{g_i,\bm p} \rangle = \lambda^2 \delta_{\bm p},~\bm p = (p_1,...,p_d),~p_i  \in \{-N_i+1,..., N_i-1\}, \nonumber \\
&~~~\left[ {\begin{array}{*{20}{c}}
	\bm Q & \bm \Phi \bm \nu \\
	(\bm \Phi \bm \nu)^H & {1}
	\end{array}} \right] \succeq 0,~\bm Q_{g_i} \succeq 0,~i = 1,...,d, \nonumber
\end{align}
where $\delta_{\bm p} = 1$ if $\bm p = (0,...,0)$ and $\delta_{\bm p} = 0$ otherwise; $\bm Q \in \mathbb{C}^{N_D \times N_D}$, $\bm Q_{g_i} \in \mathbb{C}^{{\bar N}_{D} \times {\bar N}_{D}}$; $\bm \Upsilon_{\bm p} = \bm \Upsilon_{\bm p_1} \circ \bm \Upsilon_{\bm p_2} \circ \cdots \circ \bm \Upsilon_{\bm p_d} \in \mathbb{C}^{N_D \times N_D} $ with $\circ$ being the Kronecker product and $\bm \Upsilon_{\bm p_i}$ being the $N_i\times N_i$ symmetric Toeplitz matrix generated by the $p_i$-th standard basis vector in ${\mathbb{C}^{2N_i-1}}$ according to \eqref{eq:Toeplitz};
and $\bm \Upsilon_{g_i,\bm p} \in \mathbb{C}^{{\bar N}_{D} \times {\bar N}_{D}}$ is defined with respect to $\bm \Upsilon_{\bm p}$ and the trigonometric polynomials $g_i$ in \eqref{eq:g1}, just like $\bm T_{g_i}^d$ with respect to $\bm T^d$. In particular, for $i=1,...,d$ we have $\bm \Upsilon_{g_i,\bm p} = \bm 0$ except 
\begin{align}
\bm \Upsilon_{g_i,(p_1,...,p_i,...,p_d)} =
\left\{
\begin{array}{lr}
r_1 \bm{\bar\Upsilon}_{(p_1,..., p_{i}+1,...,p_d)},~ p_i = -N_i+1, \\
\sum_{k=0}^{1} r_k \bm{\bar\Upsilon}_{(p_1,..., p_{i}+k,...,p_d)},~ p_i = -N_i+2, \\
\sum_{k=-1}^{1} r_k \bm{\bar\Upsilon}_{(p_1,..., p_{i}+k,...,p_d)},~ -N_i+3 \leq p_i \leq N_i-3 , \\
\sum_{k=-1}^{0} r_k \bm{\bar\Upsilon}_{(p_1,..., p_{i}+k,...,p_d)},~ p_i = N_i-2, \\
r_{-1} \bm{\bar\Upsilon}_{(p_1,...,p_{i}-1,...,p_d)},~ p_i = N_i-1,
\end{array}
\right. \\
p_1 \in \{-N_1+2,..., N_1-2\},...,p_{i-1} \in \{-N_{i-1}+2,..., N_{i-1}-2\}, \nonumber \\
p_{i+1} \in \{-N_{i+1}+2,..., N_{i+1}-2\},...,p_{d} \in \{-N_d+2,..., N_d-2\}, \nonumber
\end{align}
where $\bm{\bar\Upsilon}_{\bm{p}} \in \mathbb{C}^{{\bar N}_{D} \times {\bar N}_{D}}$ is defined like $\bm \Upsilon_{\bm p}$ but with $p_i \in \{-N_i+2,..., N_i-2\}$. In this way we can have
\begin{align}
\bm T^d =&~ \sum_{p_1 = -N_1+1}^{N_1-1}... \sum_{p_d = -N_d+1}^{N_d-1} \bm \Upsilon_{(p_1,...,p_d)} {\cal B}^d(p_1,...,p_d), \\
\bm T_{g_i}^d =&~ \sum_{p_1 = -N_1+1}^{N_1-1}... \sum_{p_d = -N_d+1}^{N_d-1} \bm \Upsilon_{g_i,(p_1,...,p_d)} {\cal B}^d(p_1,...,p_d), ~i=1,...,d.
\end{align}

Note that we can extend the SDP problems \eqref{eq:SDP-problem-1} and \eqref{eq:SDP-problem-2} and dual problems \eqref{eq:SDP-dual-1} and \eqref{eq:SDP-dual-2} into the multiple frequency band case according to Corollary \ref{coro: FSAN}. Optimization problems \eqref{eq:SDP-problem-1} and \eqref{eq:SDP-problem-2} are convex, while \eqref{eq:SDP-dual-1} and \eqref{eq:SDP-dual-2} are concave, hence they can be solved with standard convex solvers, e.g., CVX~\cite{boyd2004convex}. Assume that the number of the positive semidefinite constraints in \eqref{eq:SDP-problem-1}, \eqref{eq:SDP-problem-2}, \eqref{eq:SDP-dual-1} and \eqref{eq:SDP-dual-2} is $N_p$, then the complexity in each iteration is ${\cal O}(N_pN_D^6)$ if the interior point method is used. The high computational load makes it difficult to apply the MD-FS atomic norm in large problems. Hence, in the next section, we develop fast solvers for solving \eqref{eq:SDP-problem-1} and \eqref{eq:SDP-problem-2}.

\section{ADMM-based Fast Solvers}


\subsection{An ADMM-based Algorithm for Solving \eqref{eq:SDP-problem-1}}

To solve \eqref{eq:SDP-problem-1} based on the ADMM algorithm~\cite{boyd2011distributed}, we first convert it into the following optimization problem
\begin{align}
\label{eq:ADMM-problem-1-nl}
 \min_{\bm x, {\cal B}^d, {t} } &~ \frac{1}{2N_D} {\rm Tr}(\bm T^d({\cal B}^d)) + \frac{1}{2} {t} + \mathbb{I}_{\infty}(\bm \Theta \succeq 0)  + \sum_{i=1}^d \mathbb{I}_{\infty}(\bm T_{g_i}^d({\cal B}^d) \succeq 0), \\
~{\text{s.t.}}&~ \bm \Theta = \left[ {\begin{array}{*{20}{c}}
	\bm T^d({\cal B}^d) & \bm x \\
	\bm x^H & {t}
	\end{array}} \right], ~ \bm y = \bm \Phi \bm x, \nonumber
\end{align}
where $\mathbb{I}_{\infty}(\cdot)$ is an indicator function that is $0$ if the condition in the bracket is true, and infinity otherwise. Dualize the equality constraints by an augmented Lagrangian yields
\begin{align}
\label{eq:dual-Lagrangian}
{\xi}_{\rho}(\bm x, {\cal B}^d, {t}, \bm \Theta, \bm{\widetilde U}) = &~ \frac{1}{2N_D} {\rm Tr}(\bm T^d({\cal B}^d)) + \frac{1}{2} {t} + \mathbb{I}_{\infty}(\bm \Theta \succeq 0) + \sum_{i=1}^d \mathbb{I}_{\infty}(\bm T_{g_i}^d({\cal B}^d) \succeq 0)   \nonumber \\
& + \left \langle \bm{\widetilde U}, \left[ {\begin{array}{*{20}{c}} 
\bm \Theta - \left[ {\begin{array}{*{20}{c}}
	\bm T^d({\cal B}^d) & \bm x \\
	\bm x^H & {t}
	\end{array}} \right], 
& \left[ {\begin{array}{*{20}{c}}
\bm y - \bm \Phi \bm x \\
0
\end{array}} \right] 
\end{array}} \right]   \right \rangle_{\Re}   \nonumber \\
& + \frac{\rho}{2} \left \| \bm \Theta - \left[ {\begin{array}{cc}
	\bm T^d({\cal B}^d) & \bm x \\
	\bm x^H & {t}
	\end{array}} \right] \right \|_F^2 + \frac{\rho}{2} \| \bm y - \bm \Phi \bm x \|_2^2,
\end{align}
where $\rho>0$ is the penalty parameter and 
\begin{align}
\label{eq:tildeU}
\bm{\widetilde U} \triangleq 
\left[ {\begin{array}{*{20}{c}}
\bm U, &
\left[ {\begin{array}{*{20}{c}}
\bm u\\
0
\end{array}} \right] 
\end{array}} \right]  \in \mathbb{C}^{(N_D+1)\times (N_D+2)}
\end{align}
is the dual variable with $\bm U \in \mathbb{C}^{(N_D+1)\times (N_D+1)}$ and $\bm u \in \mathbb{C}^{N_D \times 1}$. The ADMM algorithm consists of following update steps~\cite{boyd2011distributed}
\begin{align}
\label{eq:ADMM-step-1-nl}
(\bm x^{q+1}, ({\cal B}^d)^{q+1}, {t}^{q+1}) =&~ \arg \min_{ \bm x, {t}, {\cal B}^d } {\xi}_{\rho}(\bm x, {\cal B}^d, {t}, \bm \Theta^{q}, \bm{\widetilde U}^{q}), \\
\label{eq:ADMM-step-2-nl}
\bm \Theta^{q+1} =&~ \arg \min_{\bm \Theta \succeq 0} {\xi}_{\rho}(\bm x^{q+1}, ({\cal B}^d)^{q+1}, {t}^{q+1}, \bm \Theta, \bm{\widetilde U}^{q}),\\
\label{eq:ADMM-step-3-nl}
\bm{\widetilde U}^{q+1} =&~ \bm{\widetilde U}^{q} + \rho 
\left[ {\begin{array}{*{20}{c}}
\bm \Theta^{q+1} - 
\left[ {\begin{array}{*{20}{c}}
	\bm T^d(({\cal B}^d)^{q+1}) & \bm x^{q+1} \\
	(\bm x^{q+1})^H & {t}^{q+1}
\end{array}} \right], &  \left[ {\begin{array}{*{20}{c}}
\bm y - \bm \Phi \bm x^{q+1} \\
0
\end{array}} \right] 
\end{array}} \right] ,
\end{align}
where the initial iteration is started by setting $\bm \Theta^{0}$ and $\bm{\widetilde U}^{0}$ as all-zero matrices, and the iteration continues until the maximum iteration number $Q$ is reached. We next give the detailed expressions of \eqref{eq:ADMM-step-1-nl}-\eqref{eq:ADMM-step-3-nl}.

\subsubsection{Exact update of $\bm x$, $t$ and intermediate update of ${\cal B}^d$ in \eqref{eq:ADMM-step-1-nl}}

The main difficulty of solving \eqref{eq:ADMM-step-1-nl} is that when updating ${\cal B}^d$, the constraints $\bm T_{g_i}^d({\cal B}^d) \succeq 0,~i=1,...,d$ need to be satisfied simultaneously. Here we first calculate its gradients by ignoring the constraints and set them to zeros to update the variables, and then project $\bm T_{g_i}^d({\cal B}^d),~i=1,...,d$ onto the semidefinite cone to approximately refine ${\cal B}^d$. 
Define
\begin{align}
\label{eq:barTheta}
\bm \Theta = \left[ {\begin{array}{cc}
	\bm{\bar\Theta} & \bm{\bar{\theta}} \\
	\bm{\bar{\theta}}^H & \Theta
	\end{array}} \right],~
\bm U = \left[ {\begin{array}{cc}
	\bm{\bar{U}} & \bm{\bar{u}} \\
	\bm{\bar{u}}^H & u
	\end{array}} \right],
\end{align}
where $\bm{\bar\Theta}$ and $\bm{\bar{U}}$ are $N_D\times N_D$ matrices, $\bm{\bar{\theta}}$ and $\bm{\bar{u}}$ are $N_D\times 1$ vectors, and $\Theta$ and $u$ are scalars. We have the following gradients (see Appendix B)
\begin{align}
\label{eq:derivative-1-nl}
\nabla_{\bm x}{\xi}_{\rho} =&~ \rho \bm \Phi^H(\bm \Phi \bm x - \bm y) - \bm \Phi^H \bm u - 2\bm{\bar{u}} + 2 \rho (\bm x - \bm{\bar{\theta}}), \\
\label{eq:derivative-2-nl}
\nabla_{{\cal B}^d(p_1,...,p_d)}{\xi}_{\rho} =&~  \frac{1}{2} \delta_{\bm p} + \beta_{(p_1,...,p_d)} \left [ \rho {\cal B}^d(p_1,...,p_d) - {\mathbb{P}}(\rho \bm{\bar\Theta} + \bm{\bar{U}})(p_1,...,p_d) \right ], \\
\label{eq:derivative-3-nl}
\nabla_{{t}}{\xi}_{\rho} = &~ \frac{1}{2} - u + \rho ({t} - \Theta),\\
\text{with}~\beta_{(p_1,...,p_d)} =&~ \prod_{i=1}^d (N_i-| p_i |),~ p_i \in \{-N_i+1,...,N_i-1\},
\end{align}
where ${\mathbb{P}}(\cdot)$ denotes an inverse operation on the $N_D\times N_D$ input $d$-level block Toeplitz matrix, that outputs a $(2N_1-1)\times \cdots \times(2N_d-1)$ $d$-way tensor. In particular, for the $d$-level block Toeplitz matrix ${\bm T^d}\in\mathbb{C}^{N_D\times N_D}$ in \eqref{eq:level-Toeplitz}, then the $(p_1,...,p_d)$-th element of ${\mathbb{P}}({\bm T}^d)$ is given by
\begin{eqnarray}
\label{eq:PUp1pd}
{\mathbb{P}}({\bm T}^d)(p_1,...,p_d) = \frac{1}{\beta_{(p_1,...,p_d)}} \sum_{m_1-n_1=p_1,...,}^{m_d-n_d=p_d} \bm T^d(m_1,n_1;...;m_d,n_d),~m_i,n_i = 1,2,...,N_i.
\end{eqnarray}

Setting the gradients in \eqref{eq:derivative-1-nl}-\eqref{eq:derivative-3-nl} to zeros, after some manipulations we can have
\begin{align}
\label{eq:update-x-nl}
\bm x^{q+1} =&~ (\rho \bm \Phi^H \bm \Phi + 2\rho \bm I_{N_D})^{-1} ( \rho \bm \Phi^H \bm y + \bm \Phi^H \bm u^{q} + 2\bm{\bar{u}}^{q} + 2 \rho \bm{\bar{\theta}}^q), \\
\label{eq:update-B-nl}
({\cal B}^d_{\text{temp}})^{q+1} = &~ {\mathbb{P}}(\bm{\bar\Theta}^{q} + \bm{\bar{U}}^{q} / \rho ) - \frac{1}{2\rho N_D}{\cal E}_{N_D}, \\
\label{eq:update-nu-nl}
{t}^{q+1} =&~ \Theta^{q} + (u^q - 1/2)/\rho,
\end{align}
where $\bm I_{N_D}$ is an $N_D \times N_D$ identity matrix and ${\cal E}_{N_D} \in\mathbb{C}^{(2N_1-1)\times \cdots \times(2N_d-1)}$ is a $d$-way all-zero tensor except for the $(N_1,N_2,...,N_d)$-th element which is 1.

\subsubsection{Approximate refinement of ${\cal B}^d$ in \eqref{eq:ADMM-step-1-nl}}

After ${\cal B}^d$ is updated by \eqref{eq:update-B-nl}, we respectively calculate $\bm T_{g_i}^d({\cal B}^d_{\text{temp}}),~i=1,...,d$ according to \eqref{eq:g1} and \eqref{dTgd} and project them onto the semidefinite cone to obtain $[\bm T_{g_i}^d({\cal B}^d_{\text{temp}})]^{\text{PSD}},~i=1,...,d$, by computing the eigendecomposition of the matrix and setting all negative eigenvalues to zero. Afterwards we refine ${\cal B}^d$ from ${\mathbb P}\left\{ [\bm T_{g_i}^d({\cal B}^d_{\text{temp}})]^{\text{PSD}} \right\},~i=1,...,d$. However, this is an underdetermined problem, i.e., there are fewer independent equations than unknowns when calculating ${\cal B}^d$ from ${\mathbb P}\left\{ [\bm T_{g_i}^d({\cal B}^d_{\text{temp}})]^{\text{PSD}} \right\}$. To that end, we refine ${\cal B}^d$ in an iterative manner based on \eqref{dTgd}. Note that the true ${\cal B}^d_{\star}$ satisfies
\begin{align}
\label{eq:truth-B}
&~~~~{\mathbb P}\left\{ [\bm T_{g_i}^d({\cal B}^d_{\star})]^{\text{PSD}} \right\}(p_1,...,p_d) = r_{-1,i}{\cal B}^d_{\star}(p_1,...,p_i+1,...,p_d) \nonumber \\
&~~~~~~~~~~~~~~~~~~~~~~~~~~~~~~~~~~~~+ r_{0,i}{\cal B}^d_{\star}(p_1,...,p_d) + r_{1,i} {\cal B}^d_{\star} (p_1,...,p_i-1,...,p_d) \\
&\Leftrightarrow \left( 1- \frac{\varrho}{1+\varrho} \right ) r_{0,i}{\cal B}^d_{\star}(p_1,...,p_d) = \frac{1}{1+\varrho}   {\mathbb P}\left\{ [\bm T_{g_i}^d({\cal B}^d_{\star})]^{\text{PSD}} \right\}(p_1,...,p_d) \nonumber \\
&~~~~~~~~~~~~~~~~~~~~ - \frac{1}{1+\varrho} \left( r_{-1,i}{\cal B}^d_{\star}(p_1,...,p_i+1,...,p_d) + r_{1,i} {\cal B}^d_{\star} (p_1,...,p_i-1,...,p_d) \right) \\
\label{eq:iteration-idea}
&\Leftrightarrow {\cal B}^d_{\star}(p_1,...,p_d) = \frac{1}{ r_{0,i}(1+\varrho)}   {\mathbb P}\left\{ [\bm T_{g_i}^d({\cal B}^d_{\star})]^{\text{PSD}} \right\}(p_1,...,p_d) \nonumber \\
&~~~ + \frac{1}{1+\varrho} \left(  \varrho{\cal B}^d_{\star}(p_1,...,p_d) - \frac{r_{-1,i}}{ r_{0,i}}{\cal B}^d_{\star}(p_1,...,p_i+1,...,p_d) - \frac{r_{1,i}}{ r_{0,i}} {\cal B}^d_{\star} (p_1,...,p_i-1,...,p_d) \right)
\end{align}
for $p_i \in\{ -N_i+2,...,N_i-2\},~i=1,...,d$, where $\varrho>1$ is a weight factor, and $r_{0,i}$, $r_{1,i}$ as well as $r_{-1,i} = r_{1,i}^*$ are defined in \eqref{eq:g1} respect to $g_i$. Based on \eqref{eq:iteration-idea}, we refine ${\cal B}^d$ by the procedure given in Algorithm~1, where $[\cdot]^{\kappa}$ denotes the variable at the $\kappa$-th inner iteration{\footnote{We name the iteration in Algorithm~1 as the inner iteration to distinguish it from the iteration of ADMM, and we use $[\cdot]^{\kappa}$ and $(\cdot)^q$ to represent the elements in the inner iteration and ADMM iteration, respectively.}}.

\begin{algorithm}[!h] \small
	\label{tab:A-refine}
	\caption{Approximate refinement of ${\cal B}^d$.}
	\begin{tabular}{lcl}
	$[{\cal B}^d_{\text{temp}}]^{0} \leftarrow ({\cal B}^d_{\text{temp}})^{q+1}$ \\
	\sf{For} $i=1$ \sf{to} $d$ \\
		\hspace{0.4cm}\sf{For} $\kappa=0$ \sf{to} $K-1$ \\
			\hspace{0.8cm}\sf{For} $p_1=2-N_1$ \sf{to} $N_1-2$ \\
			\hspace{0.8cm} $\vdots$ \\
			\hspace{0.8cm}\sf{For} $p_d=2-N_d$ \sf{to} $N_d-2$ \\
			\hspace{1.2cm} $[{\cal B}^d_{\text{temp}}]^{\kappa+1}(p_1,...,p_d) = \frac{1}{r_{0,i}(1+\varrho)}{\mathbb P}\left\{ [\bm T_{g_i}^d({\cal B}^d_{\text{temp}})]^{\text{PSD}} \right\}(p_1,...,p_d) $ \\
			\hspace{1.2cm} $~~~+\frac{1}{1+\varrho} \left[ \varrho [{\cal B}^d_{\text{temp}}]^{\kappa}(p_1,...,p_d) - \frac{r_{-1,i}}{r_{0,i}}[{\cal B}^d_{\text{temp}}]^{\kappa}(p_1,...,p_i+1,...,p_d) - \frac{r_{1,i}}{r_{0,i}}[{\cal B}^d_{\text{temp}}]^{\kappa}(p_1,...,p_i-1,...,p_d)  \right ]$ \\
			\hspace{0.8cm}\sf{End} \\
			\hspace{0.8cm} $\vdots$ \\
			\hspace{0.8cm}\sf{End} \\
		\hspace{0.4cm}\sf{End} \\
		\hspace{0.4cm} $[{\cal B}^d_{\text{temp}}]^{0} \leftarrow [{\cal B}^d_{\text{temp}}]^{K}$ \\
	\sf{End} \\
	$({\cal B}^d)^{q+1} \leftarrow [{\cal B}^d_{\text{temp}}]^{0}$ \\
	\end{tabular}
\end{algorithm}

Note that only an approximate $({\cal B}^d)^{q+1}$ can be obtained by Algorithm~1. The idea behind Algorithm~1 is that in each inner iteration, $({\cal B}^d_{\text{temp}})^{K}$ moves a small step toward the ``SDP direction'' by adding a weighted ${\mathbb P}\left\{ [\bm T_{g_i}^d({\cal B}^d_{\text{temp}})]^{\text{PSD}} \right\}$. And when $[\bm T_{g_i}^d({\cal B}^d_{\text{temp}})]^{\text{PSD}} = \bm T_{g_i}^d({\cal B}^d_{\text{temp}})$, the left-hand side of the inner iteration in Algorithm~1 equals to the right-hand side.

\subsubsection{Exact update of \eqref{eq:ADMM-step-2-nl} and \eqref{eq:ADMM-step-3-nl}} The update of $\bm \Theta^{q}$ in \eqref{eq:ADMM-step-2-nl} is also the projection onto the positive semidefinite cone 
\begin{align}
\label{eq:update-Theta}
\bm \Theta^{q+1} =&~ \arg \min_{\bm \Theta \succeq 0} \left\| \bm \Theta - \left[ {\begin{array}{cc}
	\bm T(({\cal B}^d)^{q+1}) & \bm x^{q+1} \\
	(\bm x^{q+1})^H & {t}^{q+1}
	\end{array}} \right] + \frac{\bm U^{q}}{\rho} \right \|_F^2,
\end{align}
which is also accomplished by setting all negative eigenvalues to zero. Moreover, $\bm{\widetilde U}$ can be exact updated directly by \eqref{eq:ADMM-step-3-nl}. Hence the proposed ADMM-based solver involves an approximation only when updating ${\cal B}^d$ in \eqref{eq:ADMM-step-1-nl}.

\subsection{An ADMM-based Algorithm for Solving \eqref{eq:SDP-problem-2}}

The SDP problem in \eqref{eq:SDP-problem-2} can also be solved by ADMM similarly. Converting it into the following optimization problem
\begin{align}
\label{eq:ADMM-problem-2}
 \min_{\bm x, {\cal B}^d, {t} } & \frac{1}{2} \| \bm y - \bm \Phi \bm x \|_2^2 + \frac{\lambda}{2N_D} {\rm Tr}(\bm T^d({\cal B}^d)) + \frac{\lambda}{2} {t}  + \mathbb{I}_{\infty}(\bm \Theta \succeq 0) + \sum_{i=1}^d \mathbb{I}_{\infty}(\bm T_{g_i}^d({\cal B}^d) \succeq 0) , \\
~{\text{s.t.}}&~ \bm \Theta = \left[ {\begin{array}{*{20}{c}}
	\bm T^d({\cal B}^d) & \bm x \\
	\bm x^H & {t}
	\end{array}} \right], \nonumber
\end{align}
and dualizing the equality constraint by an augmented Lagrangian yields
\begin{align}
\label{eq:dual-Lagrangian-2}
{\bar{\xi}}_{\rho}(\bm x, {\cal B}^d, {t}, \bm \Theta, \bm U) = &~ \frac{1}{2} \| \bm y - \bm \Phi \bm x \|_2^2 + \frac{\lambda}{2N_D} {\rm Tr}(\bm T^d({\cal B}^d)) + \frac{\lambda}{2} {t} + \mathbb{I}_{\infty}(\bm \Theta \succeq 0)  + \sum_{i=1}^d \mathbb{I}_{\infty}(\bm T_{g_i}^d({\cal B}^d) \succeq 0) \nonumber \\
& + \left \langle \bm U, \bm \Theta - \left[ {\begin{array}{*{20}{c}}
	\bm T^d({\cal B}^d) & \bm x \\
	\bm x^H & {t}
	\end{array}} \right] \right \rangle_{\Re} + \frac{\rho}{2} \left \| \bm \Theta - \left[ {\begin{array}{cc}
	\bm T^d({\cal B}^d) & \bm x \\
	\bm x^H & {t}
	\end{array}} \right] \right \|_F^2,
\end{align}
where $\bm U \in \mathbb{C}^{(N_D+1)\times (N_D+1)}$ is defined in \eqref{eq:barTheta}.
The ADMM algorithm consists of the following update steps~\cite{boyd2011distributed}
\begin{align}
\label{eq:ADMM-step-1}
(\bm x^{q+1}, ({\cal B}^d)^{q+1}, {t}^{q+1}) =&~ \arg \min_{ \bm x, {t}, {\cal B}^d } {\bar{\xi}}_{\rho}(\bm x, {\cal B}^d, {t}, \bm \Theta^{q}, \bm U^{q}), \\
\label{eq:ADMM-step-2}
\bm \Theta^{q+1} =&~ \arg \min_{\bm \Theta \succeq 0} {\bar{\xi}}_{\rho}(\bm x^{q+1}, ({\cal B}^d)^{q+1}, {t}^{q+1}, \bm \Theta, \bm U^{q}),\\
\label{eq:ADMM-step-3}
\bm U^{q+1} =&~ \bm U^{q} + \rho \left( \bm \Theta^{q+1} - \left[ {\begin{array}{cc}
	\bm T^d(({\cal B}^d)^{q+1}) & \bm x^{q+1} \\
	(\bm x^{q+1})^H & {t}^{q+1}
	\end{array}} \right] \right),
\end{align}
where the initial iteration is started by setting $\bm \Theta^{0}$ and $\bm U^{0}$ as all-zero matrices.

The update of $\bm \Theta$ in \eqref{eq:ADMM-step-2} can be exact computed by projecting it onto the semidefinite cone as in \eqref{eq:update-Theta}. Moreover, \eqref{eq:ADMM-step-1} can be approximatly solved using a similar procedure as that for solving \eqref{eq:ADMM-step-1-nl}. In particular, we calculate the gradients of ${\bar{\xi}}_{\rho}(\bm x, {\cal B}^d, {t}, \bm \Theta^{q}, \bm U^{q})$ as
\begin{align}
\label{eq:derivative-1}
\nabla_{\bm x}{\bar{\xi}}_{\rho} =&~ \bm \Phi^H(\bm \Phi \bm x - \bm y) - 2\bm{\bar{u}} + 2 \rho (\bm x - \bm{\bar{\theta}}), \\
\label{eq:derivative-2}
\nabla_{{\cal B}^d(p_1,...,p_d)}{\bar{\xi}}_{\rho} =&~  \frac{\lambda}{2} \delta_{\bm p} + \beta_{(p_1,...,p_d)} \left [ \rho {\cal B}^d(p_1,...,p_d) - {\mathbb{P}}(\rho \bm{\bar\Theta} + \bm{\bar{U}})(p_1,...,p_d) \right ] , \\
\label{eq:derivative-3}
\nabla_{{t}}{\bar{\xi}}_{\rho} = &~ \frac{\lambda}{2} - u + \rho ({t} - \Theta).
\end{align}
Set the gradients in \eqref{eq:derivative-1}-\eqref{eq:derivative-3} to zeros yields
\begin{align}
\label{eq:update-x}
\bm x^{q+1} =&~ (\bm \Phi^H \bm \Phi + 2\rho \bm I_{N_D})^{-1} ( \bm \Phi^H \bm y + 2\bm{\bar{u}}^{q} + 2 \rho \bm{\bar{\theta}}^q), \\
\label{eq:update-B}
({\cal B}^d_{\text{temp}})^{q+1} = &~ {\mathbb{P}}(\bm{\bar\Theta}^{q} + \bm{\bar{U}}^{q} / \rho ) - \frac{\lambda}{2\rho N_D}{\cal E}_{N_D}, \\
\label{eq:update-nu}
{t}^{q+1} =&~ \Theta^{q} + (u^q - \lambda/2)/\rho.
\end{align}
After $\bm x$, $t$ and an intermediate ${\cal B}^d$ are updated, ${\cal B}^d$ can be refined by following the similar procedure as in Algorithm~1.

Note that the ADMM algorithms also provide the dual solutions to \eqref{eq:SDP-dual-1} and \eqref{eq:SDP-dual-2}. The following proposition states that the dual solutions $\bm{\widehat \nu}$ of \eqref{eq:SDP-dual-1} and \eqref{eq:SDP-dual-2} can be respectively obtained according to $\bm{\widetilde{U}}$ in \eqref{eq:dual-Lagrangian} and $\bm{U}$ in \eqref{eq:dual-Lagrangian-2}, which is proved in Appendix C.

\begin{proposition} \label{prop:dual} Assume that the dual solutions to \eqref{eq:ADMM-problem-1-nl} and \eqref{eq:ADMM-problem-2} are respectively $\bm{\widehat{\widetilde{U}}}$ and $\bm{\widehat{U}}$, which are defined in \eqref{eq:tildeU} with $\bm{\widetilde{U}}$ and $\bm{U}$ replaced by $\bm{\widehat{\widetilde{U}}}$ and $\bm{\widehat{U}}$, respectively. If ${\rm rank}(\bm T^d) < \min_i N_i $, then for the dual solution $\bm{\widehat \nu}$ in \eqref{eq:SDP-dual-1}, we have
\begin{align}
\label{eq:lemma-dual-eq1}
\bm \Phi^H \bm{\widehat \nu} = \bm \Phi^H \bm{\widehat{u}} = - 2 \bm{\widehat{\bar{u}}},
\end{align}
where $\bm{\widehat{\bar{u}}}$ and $\bm{\widehat{u}}$ are defined in \eqref{eq:barTheta} and \eqref{eq:tildeU}, with $\bm{\bar{u}}$ and $\bm u$ replaced by $\bm{\widehat{\bar{u}}}$ and $\bm{\widehat{\bar{u}}}$, respectively. Moreover, for the dual solution $\bm{\widehat \nu}$ in \eqref{eq:SDP-dual-2}, we have
\begin{align}
\label{eq:lemma-dual-eq2}
\bm \Phi^H \bm{\widehat \nu} = - 2 \bm{\widehat{\bar{u}}}.
\end{align}
\end{proposition}

\begin{algorithm}[!h] \small
	\label{tab:A1}
	\caption{ADMM Algorithm for Solving \eqref{eq:SDP-problem-1}/\eqref{eq:SDP-problem-2}.}
	\begin{tabular}{lcl}
		Input $\bm{y}, \bm{\Phi}$, $N_i$, ${\mathbb{F}}_i$, $i=1,...,d$, $\rho$, $\varrho$, $Q$, $K$ and $\lambda$.\\
		1, Initialize $\bm \Theta^{0} = \bm 0$ and $\bm{\widetilde{U}}^{0} = \bm 0$/$\bm U^{0} = \bm 0$. \\
		2, Obtain $g_i,~i=1,...,d$ according to \eqref{eq:g1}. \\
		\sf{For} $q=0$ \sf{to} $Q$ \\
		\hspace{0.4cm} 3, Update $\bm x^{q+1}$ according to \eqref{eq:update-x-nl}/\eqref{eq:update-x}. \\
		\hspace{0.4cm} 4, Update $({\cal B}^d_{\text{temp}})^{q+1}$ according to \eqref{eq:update-B-nl}/\eqref{eq:update-B}. \\

		\hspace{0.4cm} 5, Update $t^{q+1}$ according to \eqref{eq:update-nu-nl}/\eqref{eq:update-nu}. \\
		\hspace{0.4cm} 6, Obtain $\bm T_{g_i}^d(({\cal B}^d_{\text{temp}})^{q+1})$ respect to $g_i,~i=1,...,d$ according to \eqref{dTgd}. \\
		\hspace{0.4cm} 7, Obtain ${\mathbb P}\left\{ [\bm T_{g_i}^d(({\cal B}^d_{\text{temp}})^{q+1})]^{\text{PSD}} \right\},~i=1,...,d$ by respectively \\
		\hspace{0.4cm} ~~~projecting $\bm T_{g_i}^d(({\cal B}^d_{\text{temp}})^{q+1})$ onto the semidefinite cone and using \eqref{eq:PUp1pd}. \\
		\hspace{0.4cm} 8, Update $({\cal B}^d)^{q+1}$ according to Algorithm~1. \\
		\hspace{0.4cm} 9, Update $\bm{\Theta}^{q+1}$ according to \eqref{eq:update-Theta}.\\
		\hspace{0.4cm} 10, Update $\bm{\widetilde{U}}^{q+1}$/$\bm U^{q+1}$ according to \eqref{eq:ADMM-step-3-nl}/\eqref{eq:ADMM-step-3}.\\
		\sf{End} \\
		11, Obtain the dual solution $\bm{\widehat\nu}$ from \eqref{eq:lemma-dual-eq1}/\eqref{eq:lemma-dual-eq2}. \\
		12, Determine $\bm{\widehat f}_{\ell},~\ell = 1,...,\widehat{r}$ by solving $|\langle \bm{\widehat\nu}, \bm \Phi \bm a(\bm f) \rangle |^2 = 1 $ or \\
		~~~$|\langle \bm{\widehat\nu}, \bm \Phi \bm a(\bm f) \rangle |^2 = \lambda $, where $f_1,...,f_d$ are searched over a set of grids. \\
		13, Estimate the complex gain $\bm{\widehat\sigma}$ via the least-squares method. \\ 
		\midrule
		Return $\bm{\widehat x}$, $\bm{\widehat f}_{\ell},~\ell = 1,...,\widehat{r}$ and $\bm{\widehat \sigma}$.\\
	\end{tabular}
\end{algorithm}

The main computational load of ADMM-based solvers is the eigendecomposition in updating $\bm T_{g_i}^d$ and $\bm \Theta$, whose complexity is ${\cal{O}}(N_D^3)$. Hence, the computational complexity of the ADMM-based solvers is ${\cal{O}}(N_pN_D^3)$, which is significantly faster than the CVX solver and is more suitable for real-time implementation. Finally we summarize the proposed ADMM-based solvers for solving \eqref{eq:SDP-problem-1} and \eqref{eq:SDP-problem-2} in Algorithm~2.

\section{Numerical Simulations}

We present numerical examples in this section for a data matrix of size $N_1\times N_2$. In the simulations, we set $N_1 = N_2 = 8$ unless otherwise stated, the coefficient of each frequency is generated with fixed unit magnitude and random phase, and the frequency pairs are randomly generated in $[0.3,0.4)\times[0.5,0.6)$, i.e., $f_{L,1} = 0.3$, $f_{H,1} = 0.4$, $f_{L,2} = 0.5$ and $f_{H,2} = 0.6$. We consider two types of ${\cal P}$ in \eqref{eq:y-1}: the first is a sampling matrix with randomly chosen $N_s$ elements that are equal to 1 and the rest elements equal to 0, which is used to evaluate the performance of the proposed methods in noiseless condition; the second type is the normalized data matrix (we set ${\cal P}$ as the all-one matrix for simplicity), which is used to evaluate the performance of the proposed methods in noisy condition. 
Then the SNR according to \eqref{eq:y-1} is defined as
\begin{align}
\text{SNR} = \frac{\mathbb{E}\{| \sum_{\ell=1}^{r} \sigma_{\ell} e^{i2\pi k_1 f_{1,\ell}} \cdots  e^{i2\pi k_d f_{d,\ell}}  |^2 \} }{ \bar\sigma_w^2 } = \frac{r}{\bar\sigma_w^2},
\end{align}
where $\bar\sigma_w^2$ is the variance of the Gaussian noise samples in \eqref{eq:y-1}.

The traditional 2D AN based method in \cite{chi2015compressive} is used as the baseline for comparison, which can be converted into similar convex SDP problems as in \eqref{eq:SDP-problem-1} and \eqref{eq:SDP-problem-2} but without the constraints $\bm T_{g_i}^d({\cal B}^d) \succeq 0,~i=1,...,d$. Moreover, the ADMM-based solvers, which are the same with the FS-ADMM solvers but without the refinement steps 6-8 in Algorithm~2, are also used for comparisons.
In the subsequent simulations, the weight factors in \eqref{eq:AN-problem-2} and the baseline comparison algorithms are set as $\lambda = \bar\sigma_w \sqrt{ 2 \log{(N_D)} }$. The penalty parameters in \eqref{eq:dual-Lagrangian} and \eqref{eq:dual-Lagrangian-2} are set as $\rho = 0.05$. The maximum iteration numbers of ADMM for noiseless condition algorithm and noisy condition algorithm are respectively set as $I = 2000$ and $I = 1000$. For noiseless condition, the weight factor and the maximum iteration number in Algorithm~1 are respectively set as $\varrho = 9$ and $K = 10$. And for noisy condition they are respectively set as $\varrho = 3$ and $K = 20$.

\begin{figure*}[!htb]
	\centering
		
	\subfloat[][]{\includegraphics[width=2.3in]{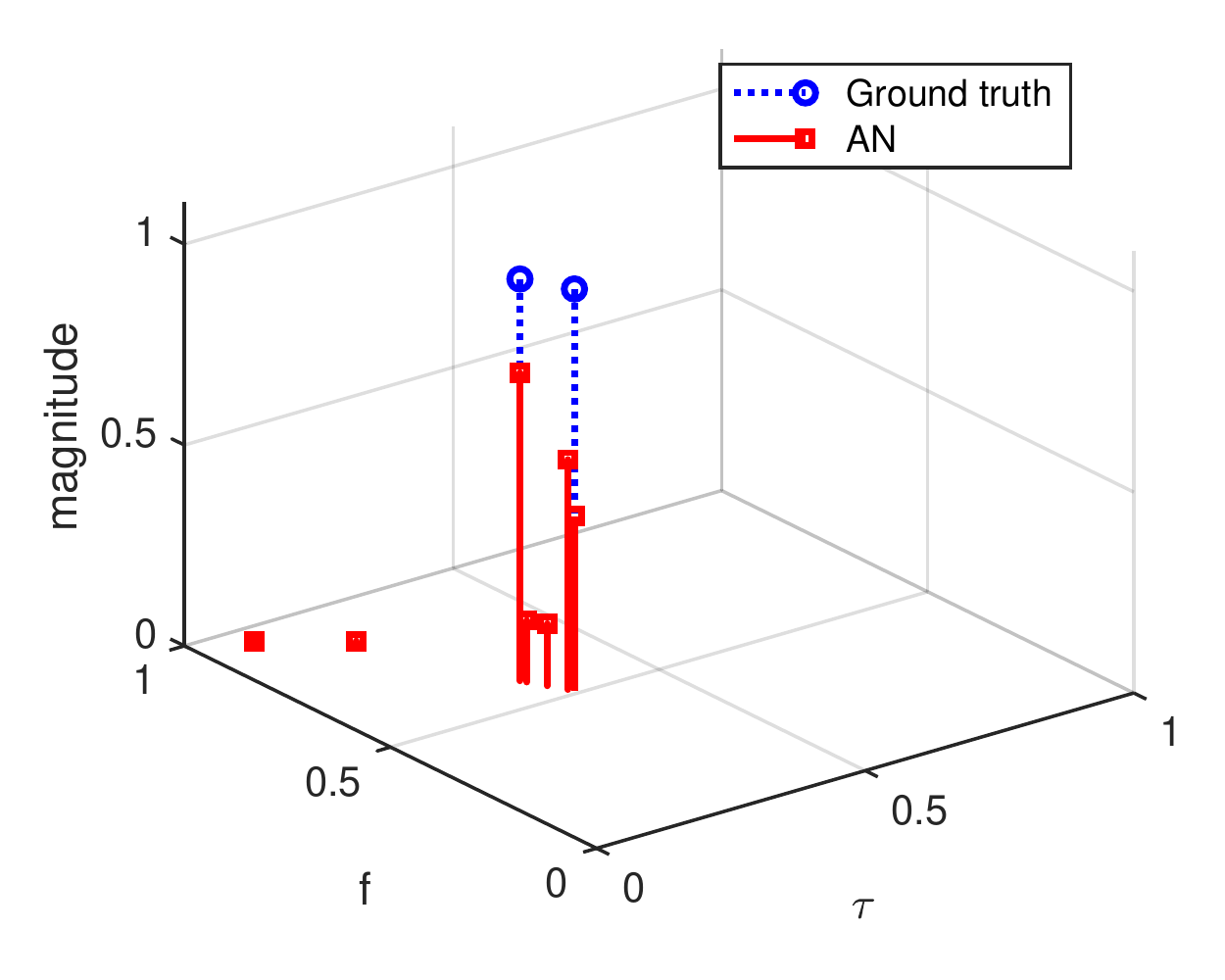}}
	\subfloat[][]{\includegraphics[width=2.3in]{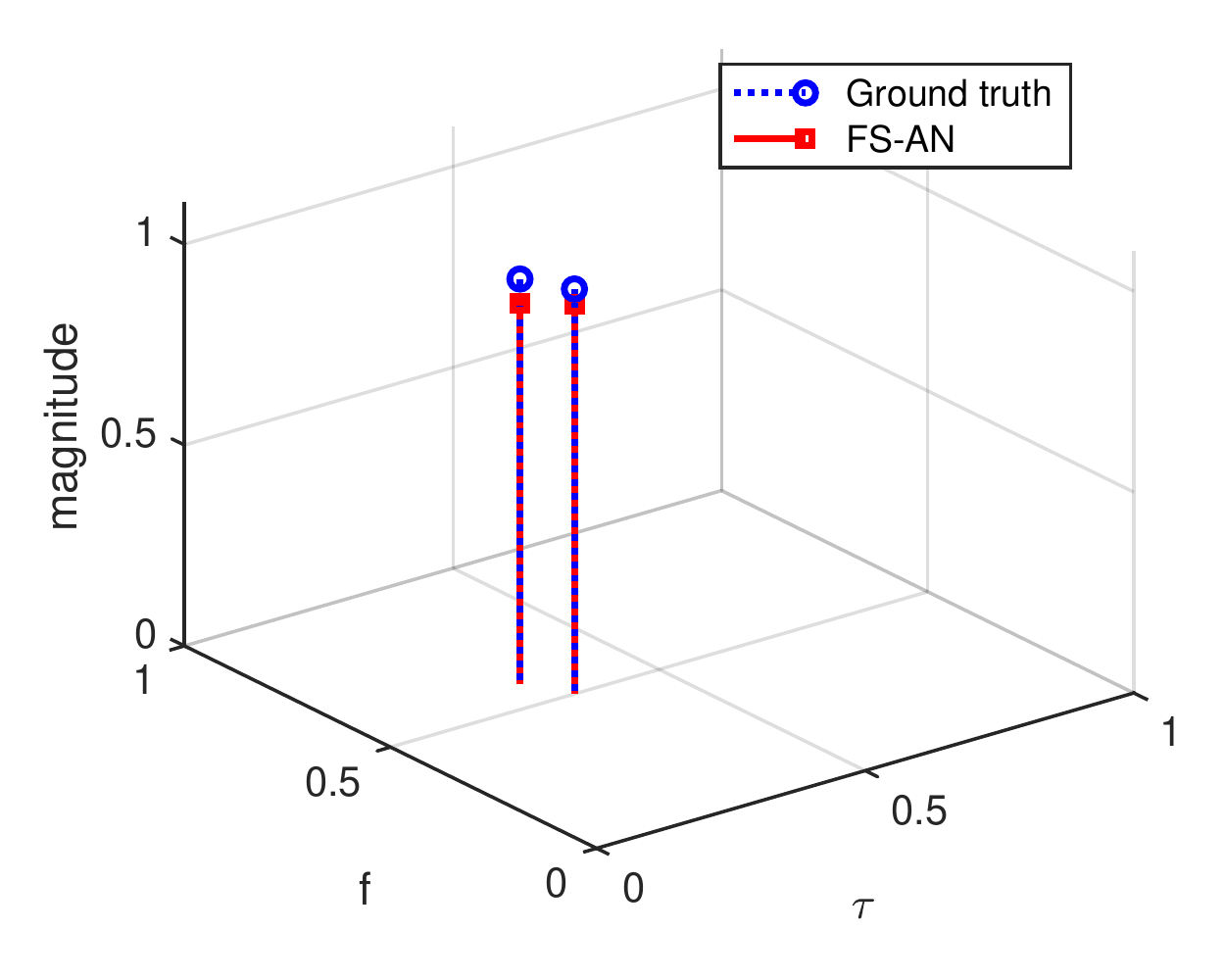}}
	\subfloat[][]{\includegraphics[width=2.3in]{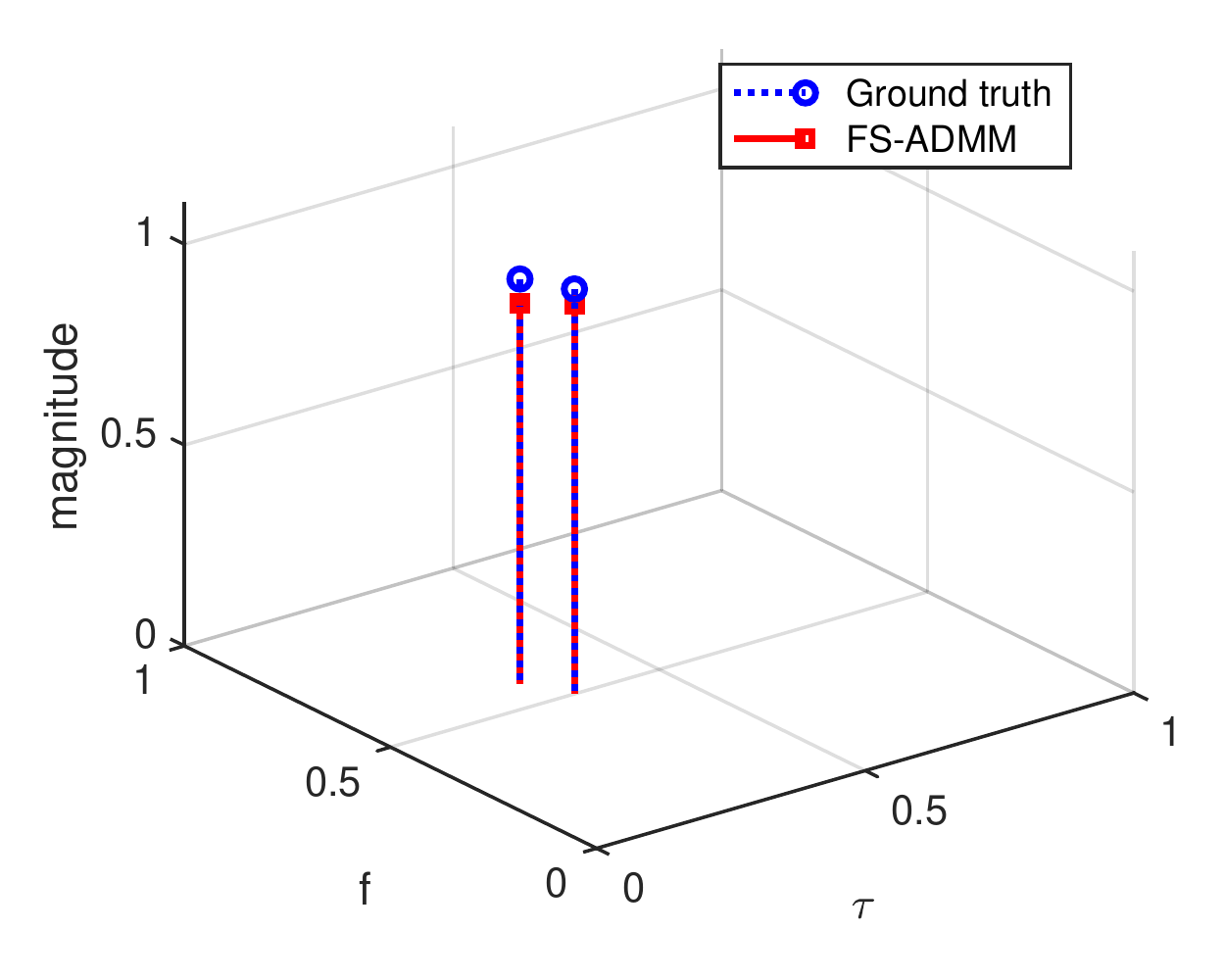}}
	
	\caption{Frequency estimation results under noisy condition with $\text{SNR} = 8~\text{dB}$. (a) AN; (b) FS-AN; (c) FS-ADMM.}
	\label{figure:frequency-results}
\end{figure*}

We first present an example when $r=2$ with $\bm f_1 = [0.35,0.51]^T$ and $\bm f_2 = [0.31,0.59]^T$ to demonstrate the effectiveness of the proposed methods in frequency estimation. In Fig.~\ref{figure:frequency-results}, the frequency estimation results of the AN, FS-AN and FS-ADMM methods in noisy condition are presented. The SNR is set as 8 dB and the 2D-MUSIC~\cite{zheng2017super,berger2010signal} is used to localize the frequencies after $\bm{\widehat{x}}$ is available. We can see that with the prior knowledge of frequency ranges, the FS-AN and FS-ADMM methods still work well under very low SNR condition, while the traditional AN method suffers from dramatic performance degradation. The dual polynomials of the FS-AN and FS-ADMM in noiseless condition (with $N_s = 40$) and noisy condition (with $\text{SNR} = 15~\text{dB}$) are shown in Fig.~\ref{figure:dual-polynomial}, where we can see that the frequencies can be determined by finding $|Q(\widehat f_1,\widehat f_2)| = 1$ and $|Q(\widehat f_1,\widehat f_2)| = \lambda$, respectively. Moreover, the dual variables $\bm{\widehat\nu}$ in Fig.~\ref{figure:dual-polynomial}(b) and (d) (provided by \eqref{eq:lemma-dual-eq1} and \eqref{eq:lemma-dual-eq2}, respectively) are identical to that in Fig.~\ref{figure:dual-polynomial}(a) and (c) (provided by solving \eqref{eq:SDP-dual-1} and \eqref{eq:SDP-dual-2}, respectively).

\begin{figure*}[!htb]
	\centering
		
	\subfloat[][]{\includegraphics[width=2.3in]{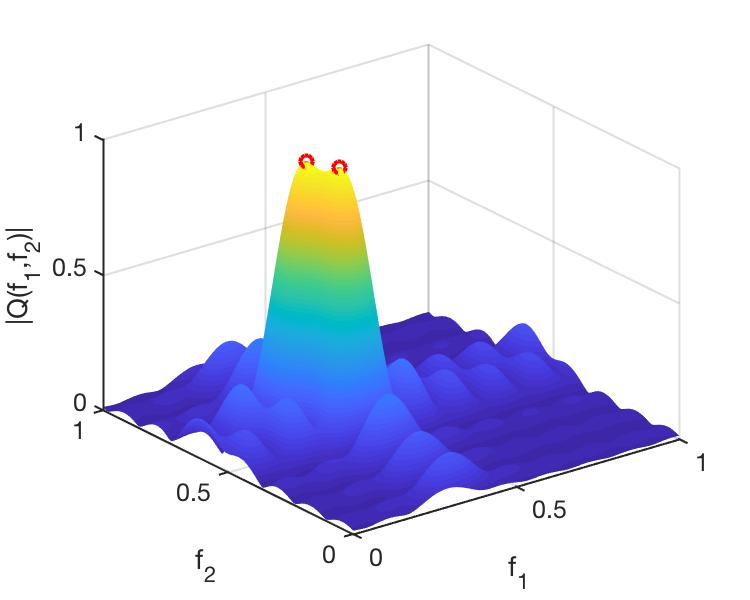}}
	\subfloat[][]{\includegraphics[width=2.3in]{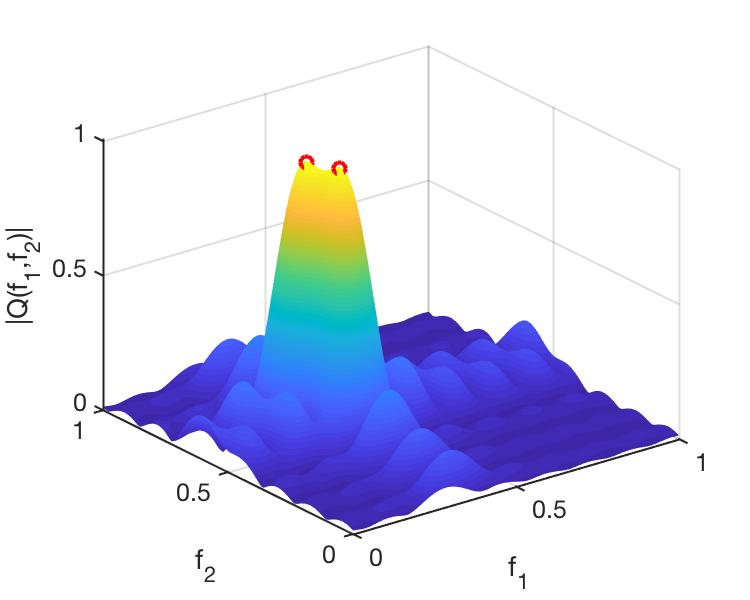}}
	
	\subfloat[][]{\includegraphics[width=2.3in]{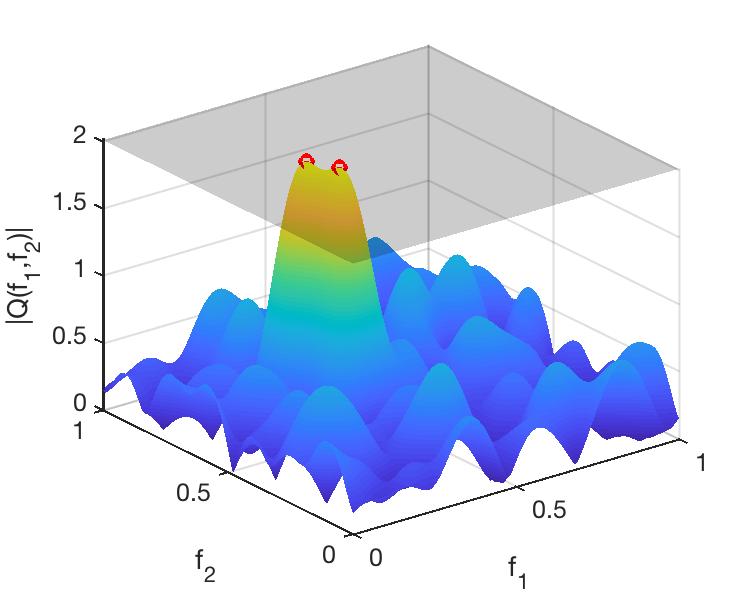}}
	\subfloat[][]{\includegraphics[width=2.3in]{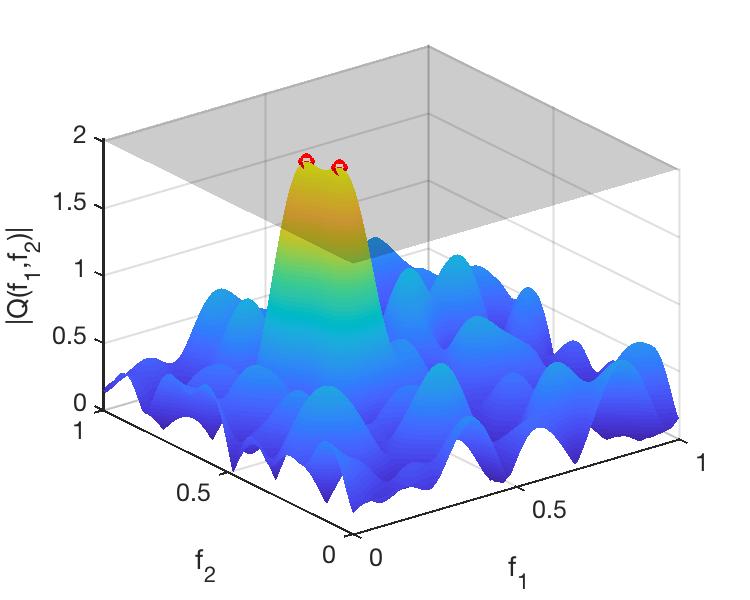}}
	
	\caption{Plots of dual polynomials in noiseless and noisy conditions. (a) FS-AN, noiseless; (b) FS-ADMM, noiseless; (c) FS-AN; noisy; (d) FS-ADMM, noisy. Dark blue represents small values while dark yellow represents large values, the ground truths are marked with red circles. The value of $\lambda = 2.0119$ is marked as transparent black planes in (c) and (d).}
	\label{figure:dual-polynomial}
\end{figure*}

\begin{figure*}[!htb]
	\centering
		
	\subfloat{\includegraphics[width=3.2in]{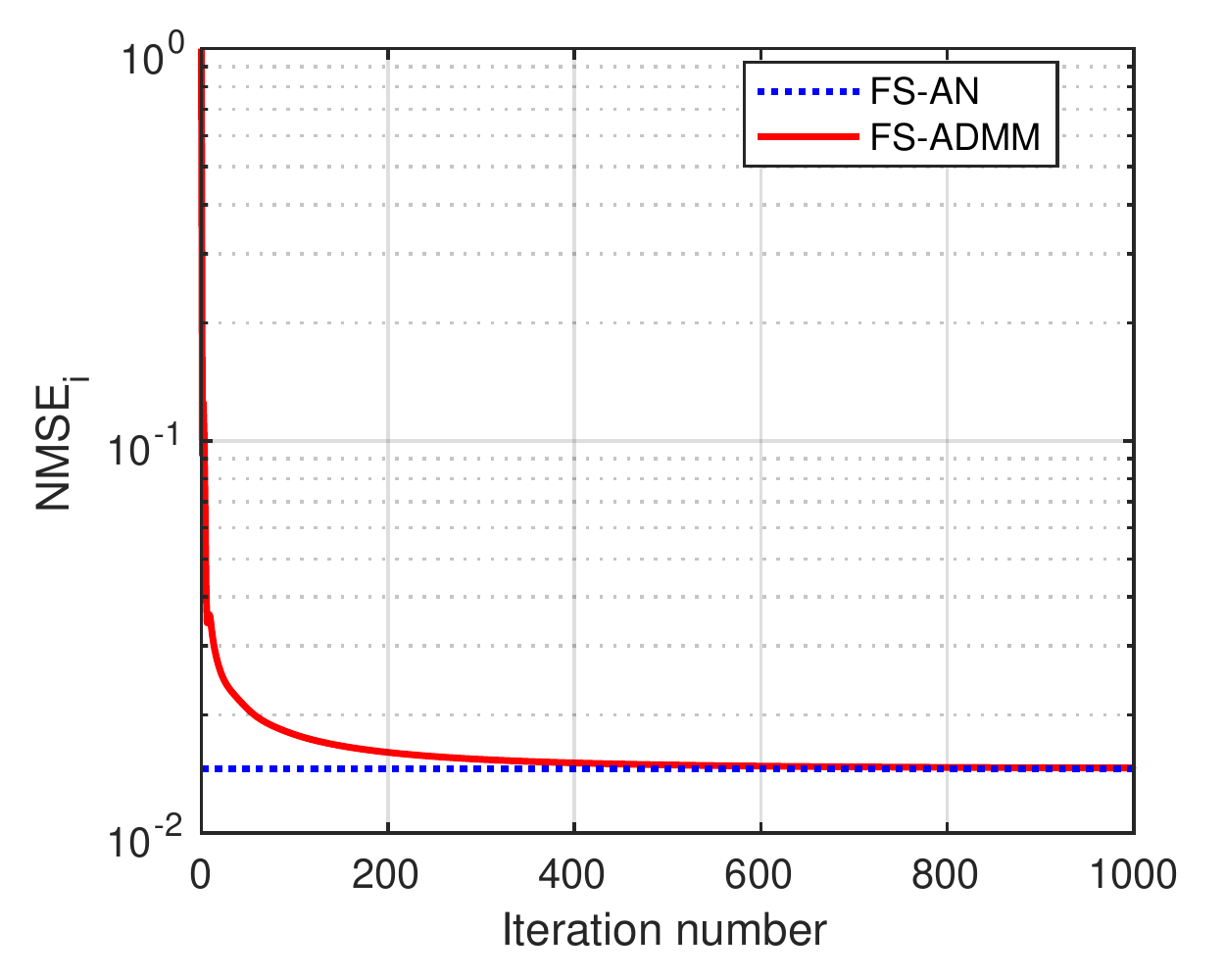}}
	
	\caption{Convergence behaviors of the proposed FS-ADMM method and FS-AN method.}
	\label{figure:convergence}
\end{figure*}

An example of the convergence behavior of the proposed methods is shown in Fig.~\ref{figure:convergence}.  The SNR is set as 20 dB. The normalized mean-squared-error (NMSE) $\|\bm{\widehat x} - \bm x\|_2 / \|\bm x\|_2$ in each ADMM iteration is calculated, we can see that the $\text{NMSE}_{i}$ of the FS-ADMM method is close to the NMSE of the FS-AN method after 1000 iterations. The running times of the proposed methods when $N_1 = N_2 = 8$, $N_1 = N_2 = 12$ and $N_1 = N_2 = 16$ are given in Table~\ref{tab:runtime}. The simulations were carried out on an Intel Xeon desktop computer with a 3.5 GHz CPU and 24 GB of RAM. We can see that the FS-ADMM method is much faster than the FS-AN method, especially for large problems.

\begin{table}[!htb]
\caption{Running time comparison}
\label{tab:runtime}
\centering
\begin{tabular}{cccc}  
\toprule
Methods  & $8\times 8$ & $12\times 12$ & $16\times 16$ \\
\midrule
FS-AN    & 41.52s & 660.57s & 6821.43s\\
FS-ADMM     & 8.82s & 28.34s & 67.66s \\
\bottomrule
\end{tabular}
\end{table}

\begin{figure*}[!htb]
	\centering
		
	\subfloat[][]{\includegraphics[width=2.3in]{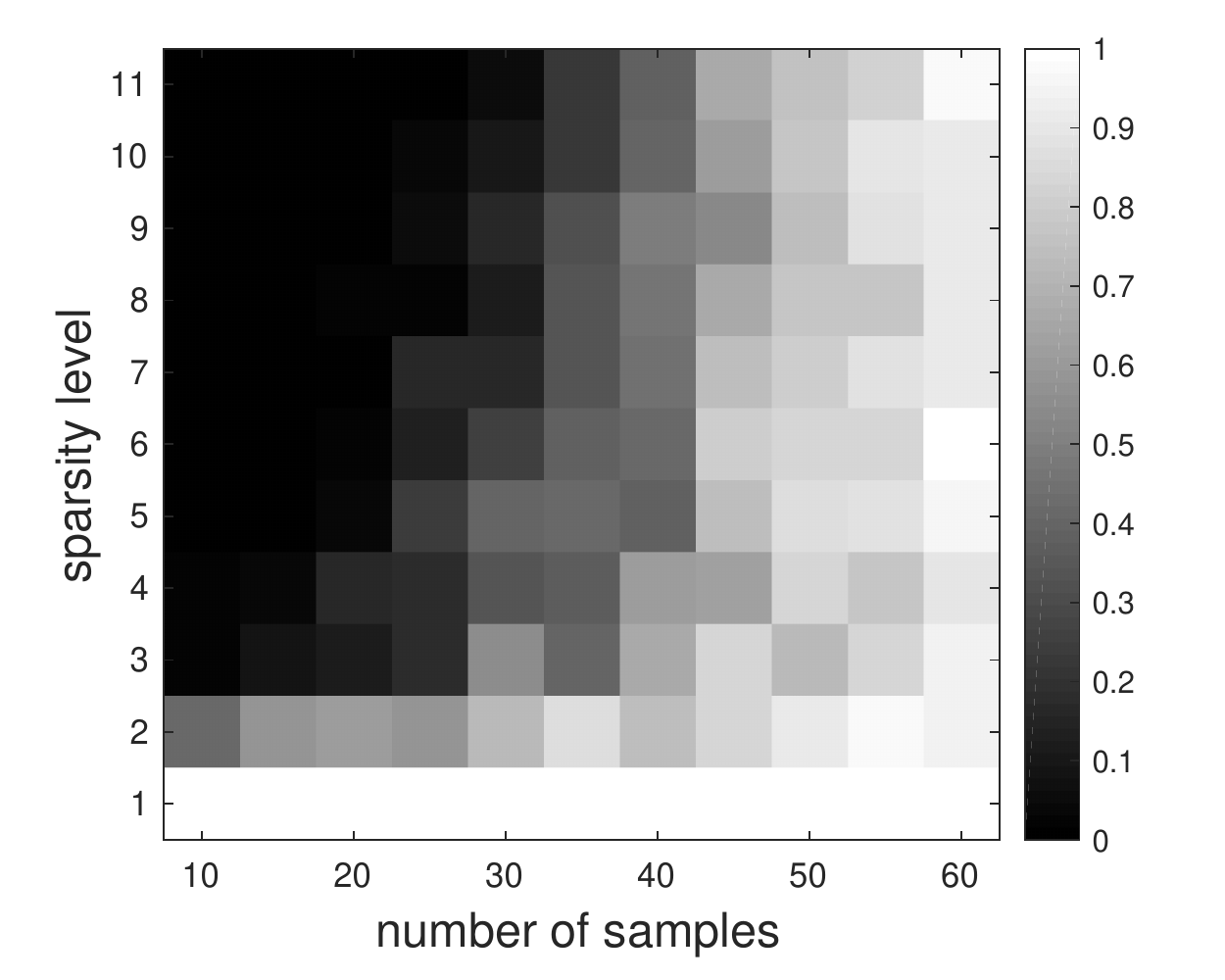}}
	\subfloat[][]{\includegraphics[width=2.3in]{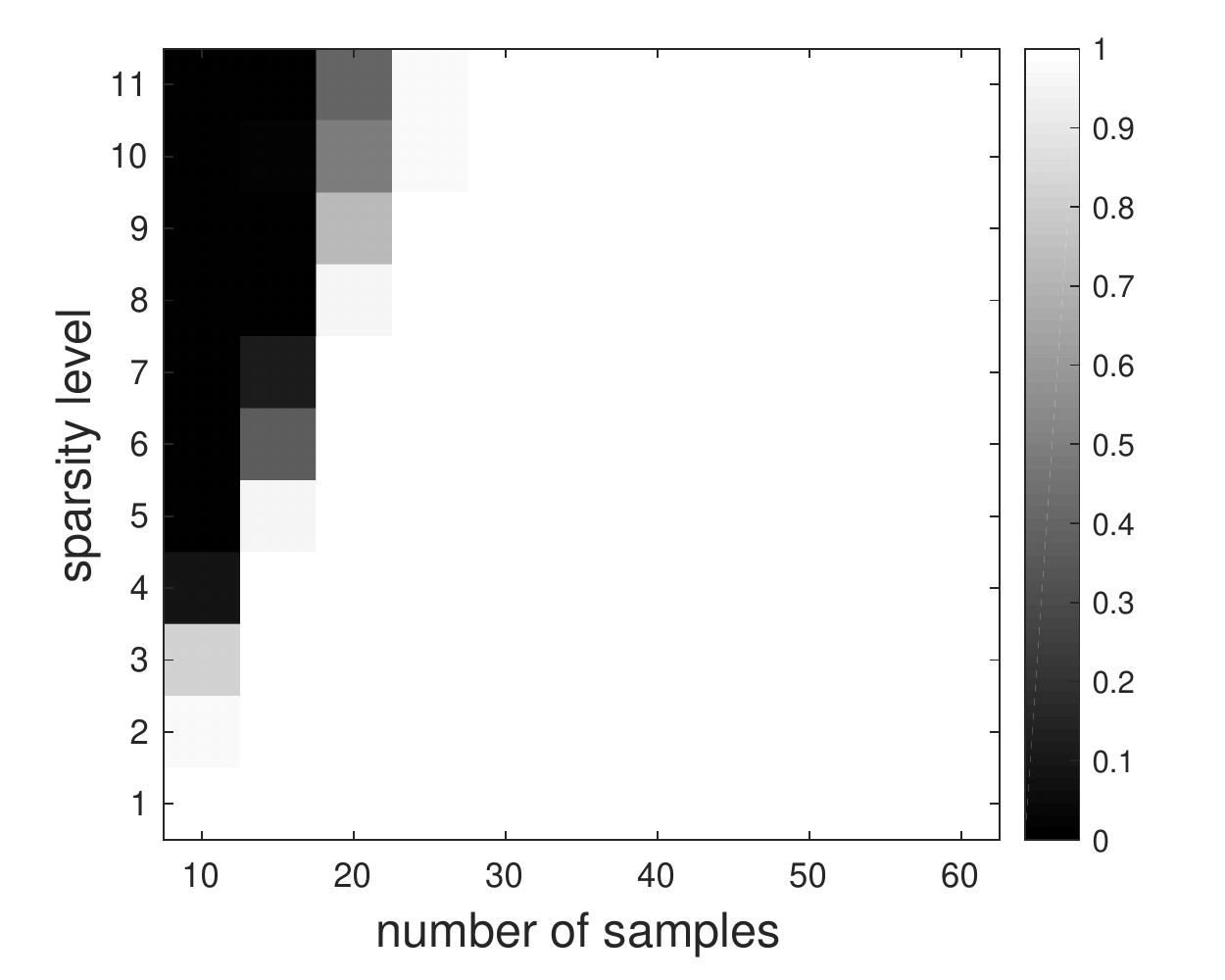}}
	\subfloat[][]{\includegraphics[width=2.3in]{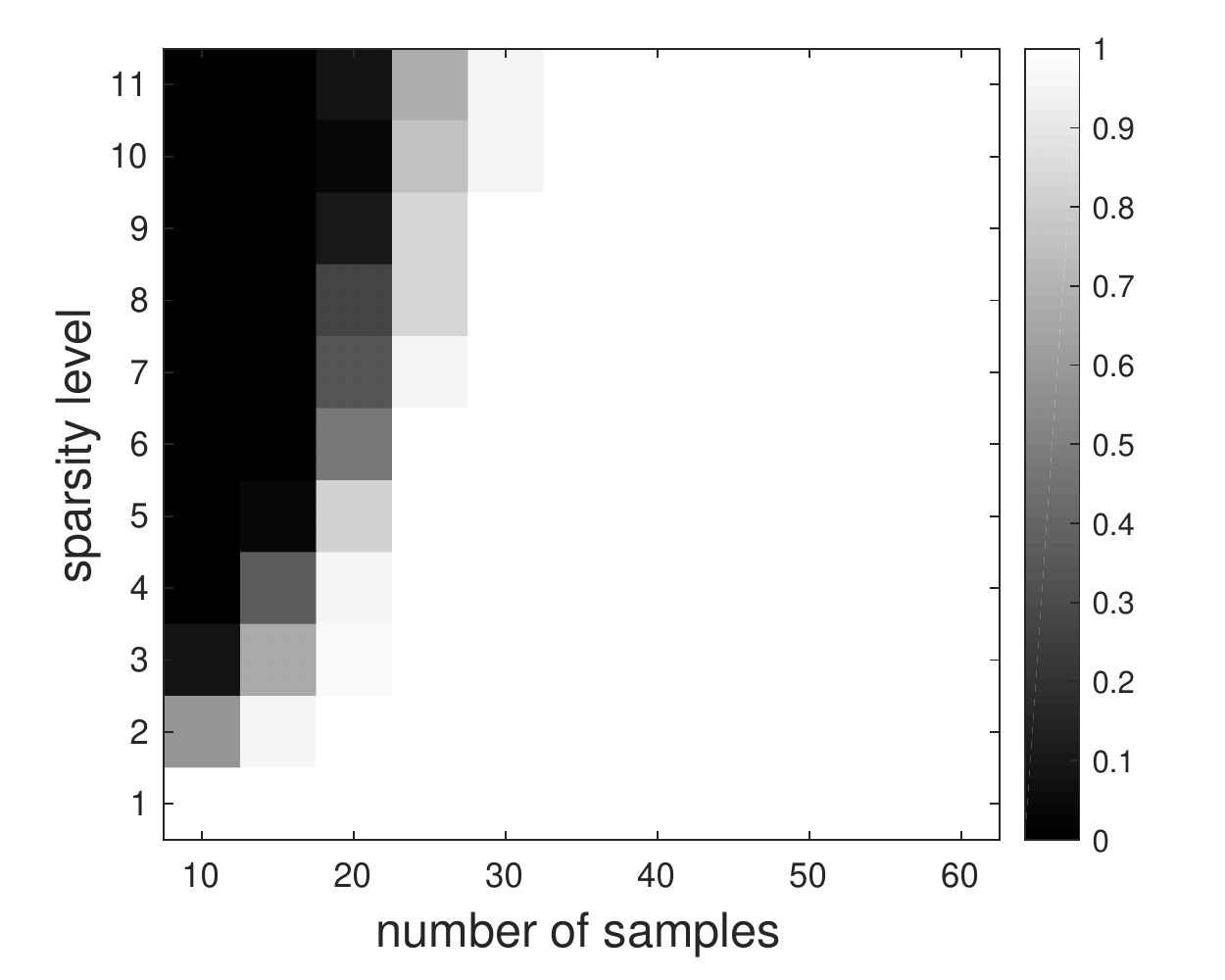}}
	
	\subfloat[][]{\includegraphics[width=2.3in]{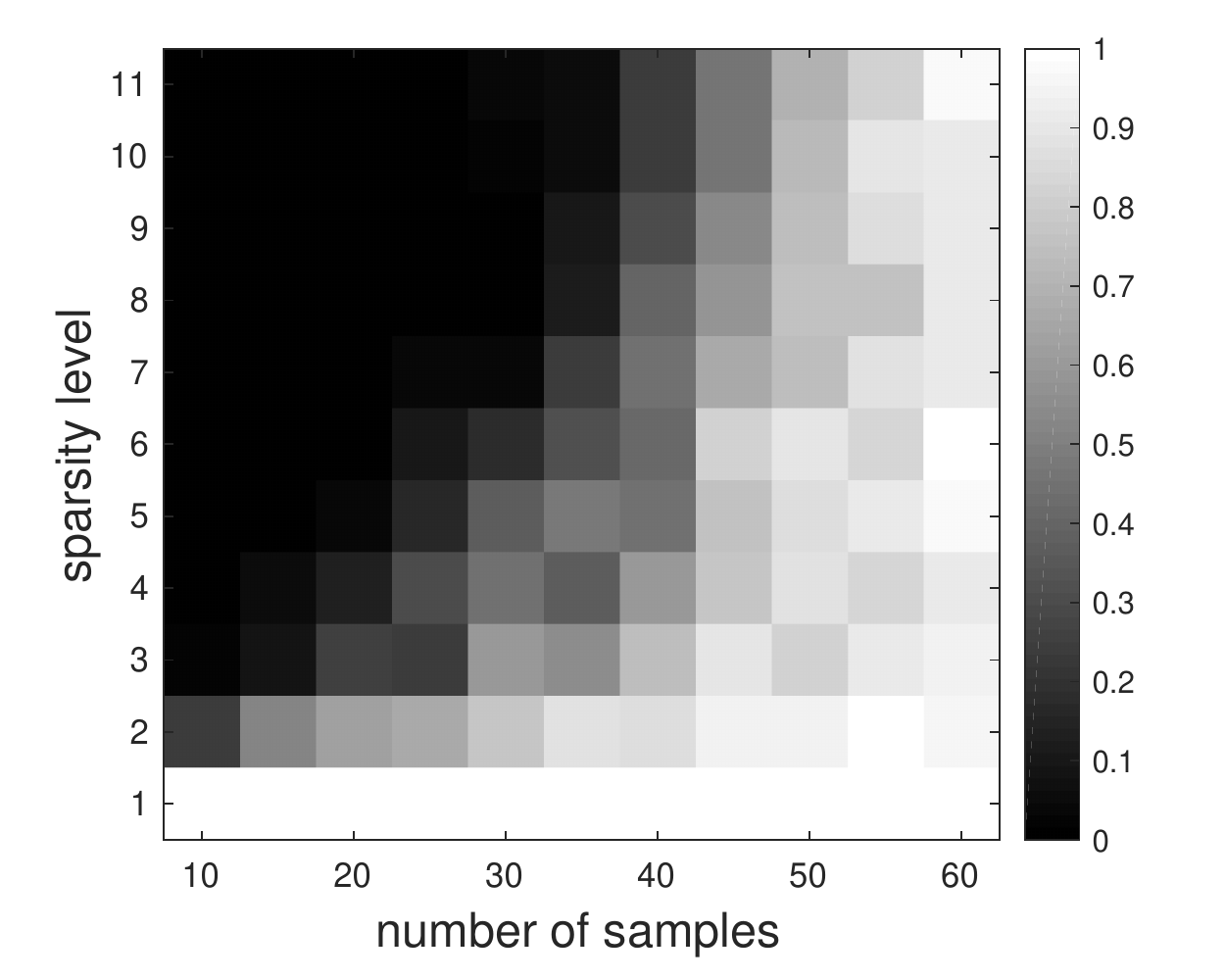}}
	\subfloat[][]{\includegraphics[width=2.3in]{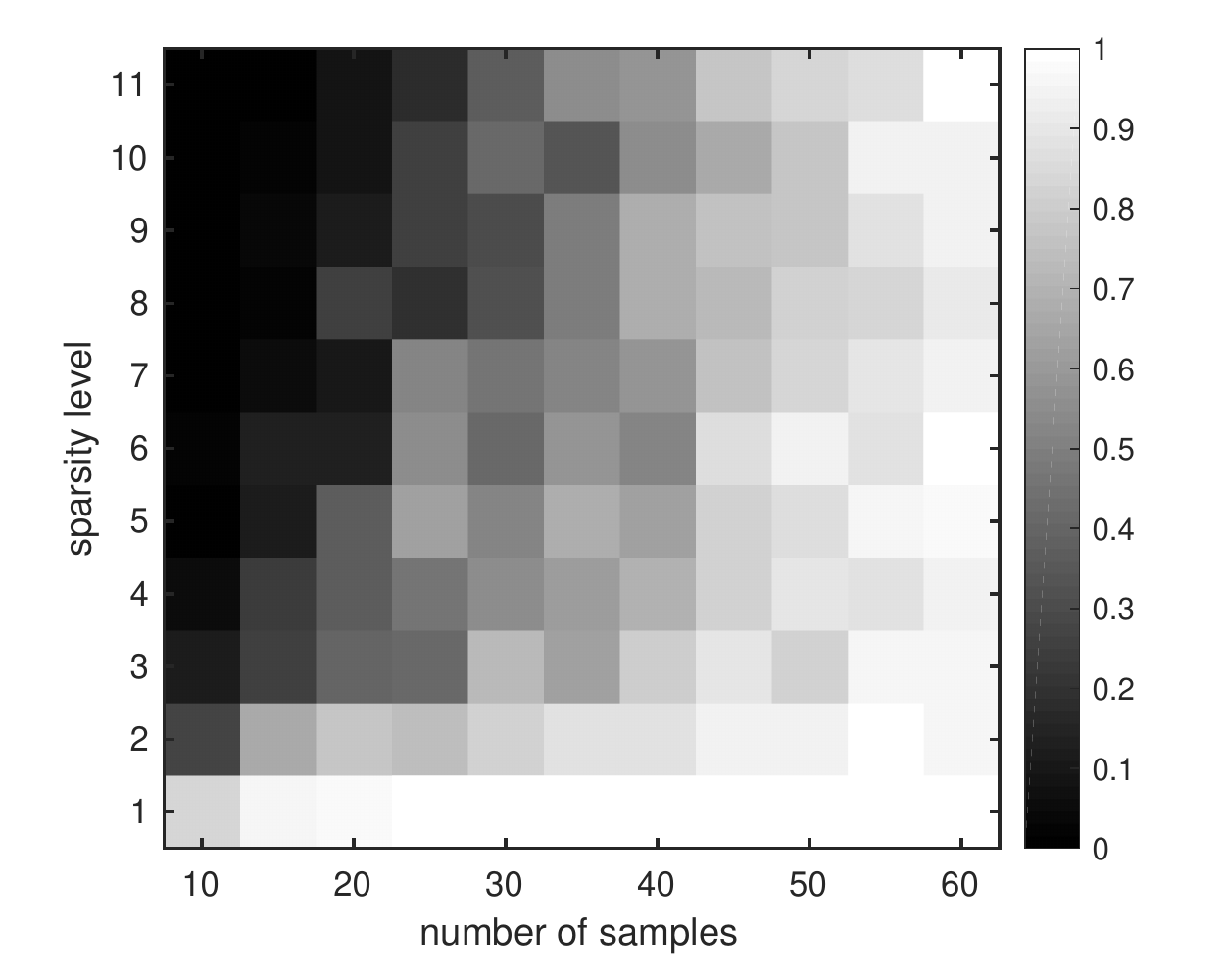}}
	\subfloat[][]{\includegraphics[width=2.3in]{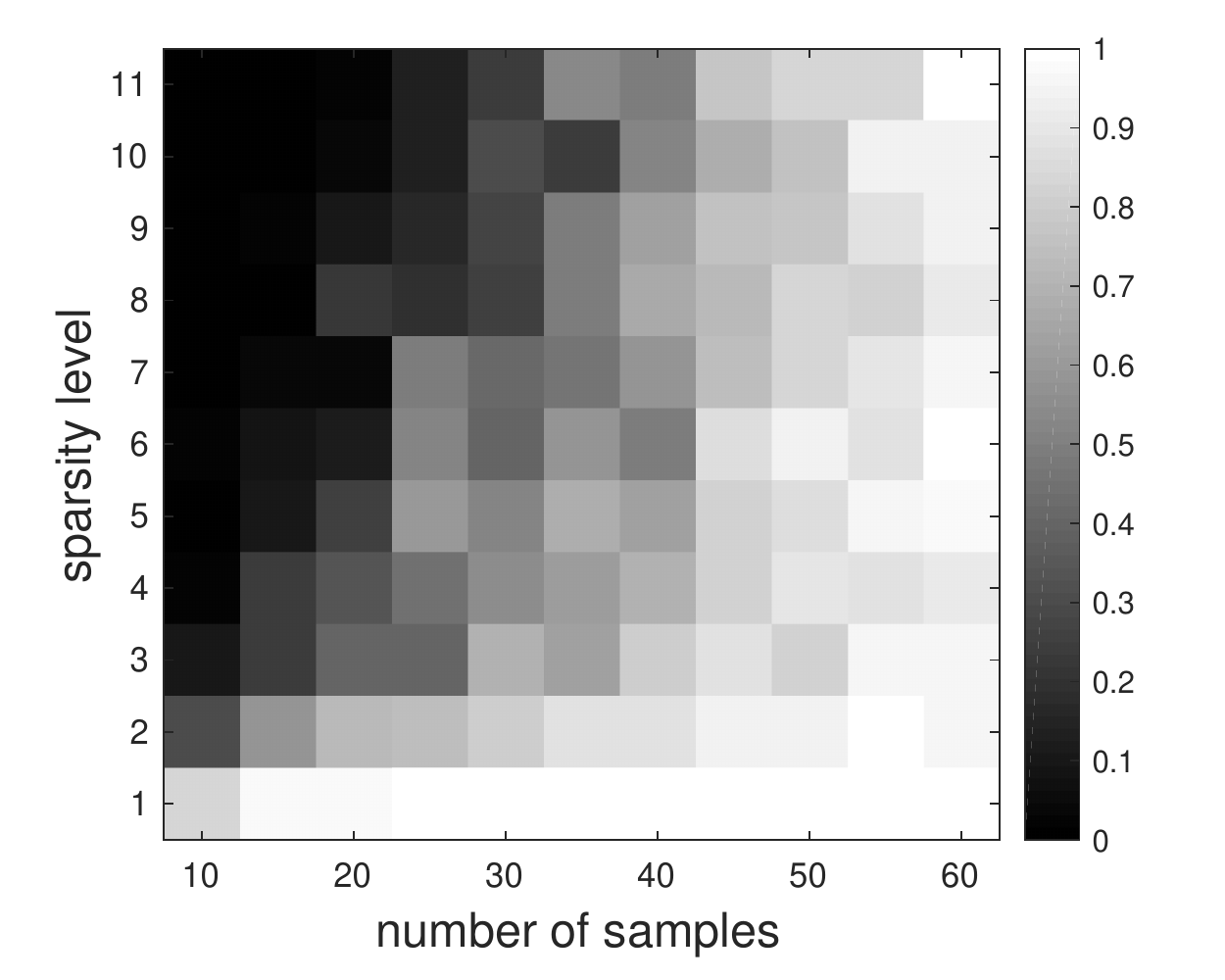}}
	
	\caption{Phase transition plots when $N_1 = N_2 = 8$. (a) AN; (b) FS-AN; (c) FS-AN with rough prior knowledge; (d) ADMM; (e) FS-ADMM; (e) FS-ADMM with rough prior knowledge. The success rate is calculated by averaging over 50 runs. The grayscale of each cell reflects the empirical rate of success.}
	\label{figure:phase-transition}
\end{figure*}

We next examine the phase transition of the proposed methods under noiseless condition. For each sample size $N_s$ and sparsity level $r$, we run 50 experiments. In each experiments, the recovery was considered successful if the NMSE of $\bm{\widehat{x}}$ is smaller than $10^{-5}$ and $10^{-3}$ for the CVX solver and ADMM-based solver, respectively. Since in noiseless condition, the CVX solver is nearly optimal and the ADMM-based approaches suffer from performance degradation due to finite iteration number. Fig.~\ref{figure:phase-transition} shows the success rates for each number of samples and $r$, with the grayscale of each cell reflecting the empirical rate of success. Sometimes only rough ranges of the frequencies may be available. Hence we also evaluate the proposed methods with rough prior knowledge as $f_{L,1} = 0.2$, $f_{H,1} = 0.4$, $f_{L,2} = 0.5$ and $f_{H,2} = 0.7$. Comparing Figs.~\ref{figure:phase-transition}(b) and (e) respectively with (a) and (d), it can be seen that with the accurate prior knowledge of frequency ranges, the performance of the 2D harmonic retrieval can be significantly improved. In addition, the proposed methods work well even if the condition ${\rm rank}(\bm T^d) < \min_i N_i $ does not hold. Moreover, from Figs.~\ref{figure:phase-transition}(c) and (f) we can see that even when the prior knowledge is not very accurate, the performance of the 2D harmonic retrieval is also improved by the proposed methods. Furthermore, the more accurate is the prior knowledge, the better is the performance of the proposed methods. 

Finally, we evaluate the root-mean-squared-error (RMSE) of the 2D frequencies estimation under the noisy condition. The 2D-MUSIC algorithm is used for frequency estimation after $\bm{\widehat{x}}$ is obtained. Note that the algorithm may return a bunch of frequencies, which can be either true detections or false alarms (see Fig.~\ref{figure:frequency-results}(a)). For each estimated frequency pair $(\widehat f_{1,\ell},\widehat f_{2,\ell}),~\ell = 1,...,\widehat r$, we calculate the minimum absolute error (AE), take $f_1$ as an example, as $\text{AE}_{f_1,\ell} = \min \left( \min(\{ \widehat f_{1,\ell} - f_{1,m} \}_{m=1}^r), \Delta_{f_1} \right )$ with $ \Delta_{f_1} = 0.1$ being the range of the prior knowledge. Then, the RMSE is calculated as
\begin{align}
\text{RMSE}_{f_1/f_2} = \sqrt{  \frac{1}{N_r} \sum_{n_r = 1}^{N_r} \frac{1}{\widehat r} \sum_{\ell = 1}^{\widehat r} (\text{AE}_{f_1/f_2,\ell}^{(n_r)})^2 },
\end{align}
where $N_r$ denotes the number of runs and $\text{AE}_{f_1/f_2,\ell}^{(n_r)}$ denotes the minimum AE of the $\ell$-th estimate in the $n_r$-th run. Fig.~\ref{figure:snr} shows the mean RMSE over $\text{RMSE}_{f_1}$ and $\text{RMSE}_{f_2}$, we can see that the frequency estimation performance is greatly improved when the prior knowledge is given, especially when the SNR is low. Note that, when the SNR is below 8 dB, the FS-ADMM method has performance degradation compared with the FS-AN method. This is because when the noise is very strong, the accuracy of the approximate refinement in Algorithm~1 will be affected. However, the FS-ADMM method is still significantly better than AN and ADMM methods, and it has moderate computational complexity and is suitable for real-time implementation.

\begin{figure*}[!htb]
	\centering
		
	\subfloat{\includegraphics[width=3.2in]{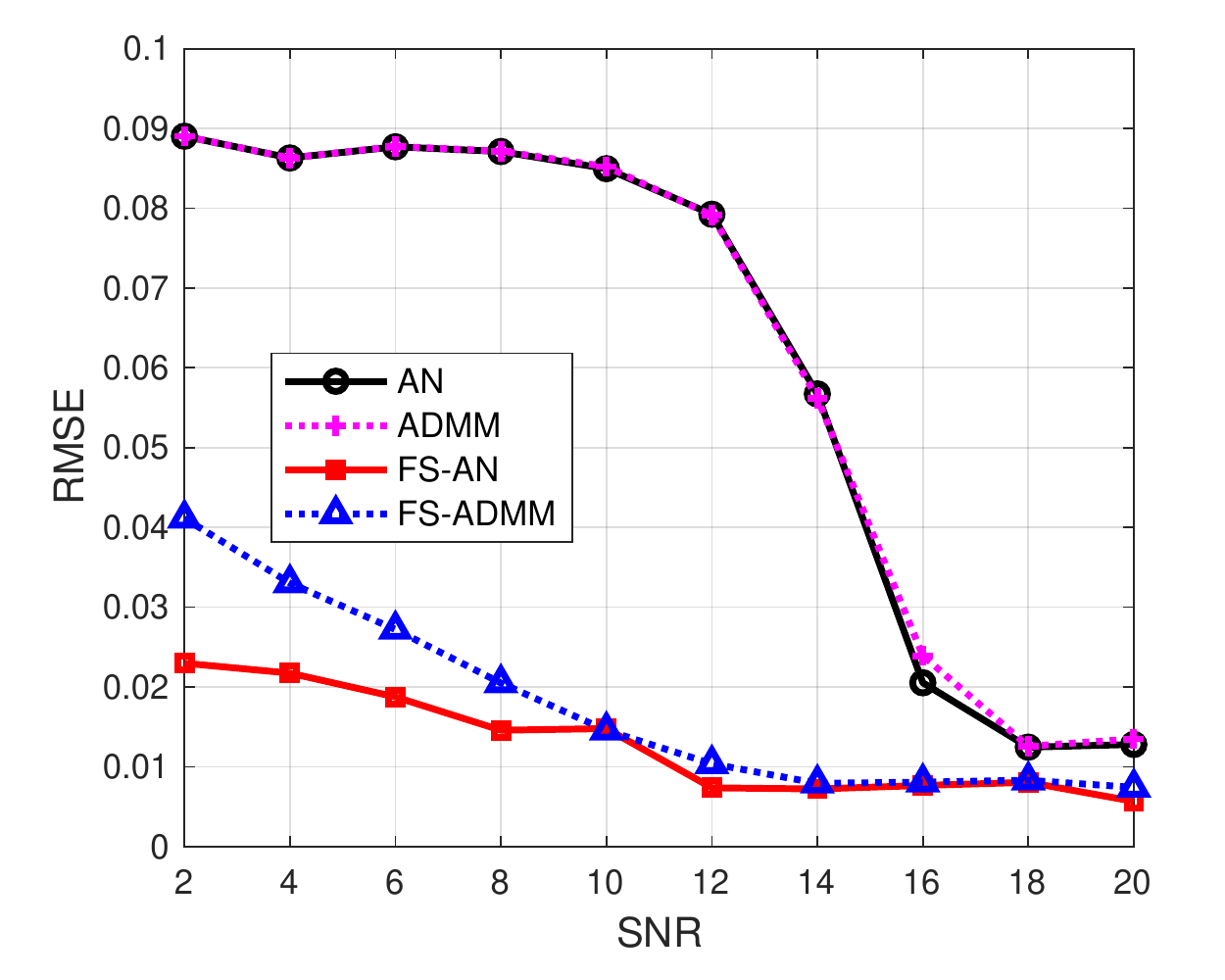}}
	
	\caption{Comparison of RMSE performances. The RMSE is calculated by averaging over 20 runs.}
	\label{figure:snr}
\end{figure*}

\section{Conclusions}

In this paper, we have treated the estimation of the frequency components of a $d$-way ($d\geq 2$) low-rank spectral tensors with the prior knowledge of the frequency intervals. We first formulated two FS atomic norm minimization problems and their dual problems for determining the frequencies. Then, the MD-FS Vandermonde decomposition of block Toeplitz matrices on given intervals is developed. It is shown that by using the MD-FS Vandermonde decomposition, we can convert the FS atomic norm minimization problems into semidefinite programs. ADMM-based fast algorithms are also developed to compute the solutions to the semidefinite programs, where each iteration contains a number of refinement steps to utilize the prior knowledge. Numerical results demonstrate the effectiveness of the proposed methods.


\appendix

\subsection{Proof of Lemma~\ref{lem: ptp}}

The proof follows closely the analysis in~\cite{yang2018frequency}, which for completeness is given here. From \eqref{eq:r0} we have
\begin{align}
r_{0,i} =&~ -\frac{x_{L,i}+x_{H,i}}{ \sqrt{x_{L,i}x_{H,i}} } {\rm sign}(f_{H,i}-f_{L,i}), \\
r_{1,i} =&~ \sqrt{x_{L,i}x_{H,i}} {\rm sign}(f_{H,i}-f_{L,i}),
\end{align}
where $x_{L,i} = e^{i2\pi f_{L,i}}$ and $x_{H,i} = e^{i2\pi f_{H,i}}$. Rewrite \eqref{eq:htps} as
\begin{align}
g_i(x) = \frac{1}{x\sqrt{x_{L,i} x_{H,i}}}(x-x_{L,i})(x-x_{H,i}){\rm sign}(f_{H,i}-f_{L,i})
\end{align}
for $i=1,...,d$. It is clear that each $g_i(x)$ is a Hermitian trigonometric polynomial and is real-valued on ${[0,1)}$. And $g_i(x)$ has two single roots $x_{L,i}$ and $x_{H,i}$, while $g_i(f)$ has two single roots $f_{L,i}$ and $f_{H,i}$. Moreover,
\begin{align}
g_i\left(\frac{1}{2}(f_{L,i}+f_{H,i})\right) =&~ r_{0,i} + 2\Re(r_{1,i} e^{-i\pi({f_{L,i}+f_{H,i}})}) \nonumber \\
=&~ \{ 2-2\cos[\pi(f_{L,i}-f_{H,i})] \} {\rm sign}(f_{H,i}-f_{L,i}).
\end{align}
Hence the sign of $g_i$ at $\frac{1}{2}(f_{L,i}+f_{H,i})$ is identical to the sign of $f_{H,i}-f_{L,i}$. This indicates that whenever $f_{L,i}<f_{H,i}$ or $f_{L,i}>f_{H,i}$, $g_i(f)$ is always positive on $(f_{L,i},f_{H,i})$ and negative on $(f_{H,i},f_{L,i})$, which completes the proof.

\subsection{Derivations of \eqref{eq:derivative-1-nl}-\eqref{eq:derivative-3-nl}}

	
Rewrite \eqref{eq:dual-Lagrangian} by ignoring the SDP constraints as
\begin{align}
\label{eq:dual-Lagrangian-re}
{\xi}_{\rho}(\bm x, {\cal B}^d, {t}, \bm \Theta, \bm{\widetilde U}) = &~ \frac{1}{2N_D} {\rm Tr}(\bm T^d({\cal B}^d)) + \frac{1}{2} {t}
+ \left\langle  \bm U , \bm{\bar\Theta} - \bm T^d({\cal B}^d) \right\rangle_{\Re}
 \nonumber \\
& + 2  \left\langle \bm{\bar{u}} , \bm{\bar{\theta}} - \bm x \right \rangle_{\Re} + \left\langle u,\Theta - t \right\rangle_{\Re} +  \left\langle  \bm u , \bm y - \bm \Phi \bm x \right \rangle_{\Re}  \nonumber \\
&+ \frac{\rho}{2} \left \| \bm{\bar\Theta} - \bm T^d({\cal B}^d) \right \|_F^2 + \rho \left \| \bm{\bar{\theta}} - \bm x \right \|_2^2 + \frac{\rho}{2}\|\Theta-t\|_2^2 + \frac{\rho}{2} \left \| \bm y - \bm \Phi \bm x \right \|_2^2.
\end{align}
Note that for a real function $f$ of a complex vector $\bm x$, the complex gradient vector is given by~\cite{petersen2008matrix}
\begin{align}
\nabla_{\bm x}f(\bm x) =  \frac{\partial f(\bm x)}{\partial \bm x_{\Re}} + i \frac{\partial f(\bm x)}{\partial  \bm x_{\Im}},
\end{align}
where $\bm x_{\Re}$ and $\bm x_{\Im}$ respectively denote the real and imaginary parts of $\bm x$, i.e., $\bm x = \bm x_{\Re} + i  \bm x_{\Im}$. Taking the gradient of $\left\langle  \bm u , \bm y - \bm \Phi \bm x \right \rangle_{\Re}$ as an example we have
\begin{align}
&~\nabla_{\bm x} \left\langle  \bm u , \bm y - \bm \Phi \bm x \right \rangle_{\Re} = - \nabla_{\bm x} \Re( \bm x^H \bm \Phi^H \bm u) \nonumber \\
=&~ - \nabla_{\bm x} ( \bm x_{\Re}^T \bm \Phi_{\Re}^T \bm u_{\Re} - \bm x_{\Im}^T \bm \Phi_{\Im}^T \bm u_{\Re} + \bm x_{\Re}^T \bm \Phi_{\Im}^T \bm u_{\Im} + \bm x_{\Im}^T \bm \Phi_{\Re}^T \bm u_{\Im} ) \nonumber \\
=&~ -\left(  \bm \Phi_{\Re}^T \bm u_{\Re} + \bm \Phi_{\Im}^T \bm u_{\Im}  -  i \bm \Phi_{\Im}^T \bm u_{\Re}   +  i \bm \Phi_{\Re}^T \bm u_{\Im}  \right) = - \bm \Phi^H \bm u.
\end{align}
Then, after some manipulations, the gradients of ${\xi}_{\rho}(\bm x, {\cal B}^d, {t}, \bm \Theta, \bm{\widetilde U})$ respect to $\bm x$ and $t$ are given by
\begin{align}
\label{eq:derivative-1-nl-ap}
\nabla_{\bm x}{\xi}_{\rho} =&~ \rho \bm \Phi^H(\bm \Phi \bm x - \bm y) - \bm \Phi^H \bm u - 2\bm{\bar{u}} + 2 \rho (\bm x - \bm{\bar{\theta}}), \\
\label{eq:derivative-3-nl-ap}
\nabla_{{t}}{\xi}_{\rho} = &~ \frac{1}{2} - u + \rho ({t} - \Theta),
\end{align}
so \eqref{eq:derivative-1-nl} and \eqref{eq:derivative-3-nl} are obtained. Next we calculate the gradient respect to ${\cal B}^d$, note that only three terms $\frac{1}{2N_D} {\rm Tr}(\bm T^d({\cal B}^d)) $, $\left\langle  \bm U , \bm{\bar\Theta} - \bm T^d({\cal B}^d) \right\rangle_{\Re}$ and $\frac{\rho}{2} \left \| \bm{\bar\Theta} - \bm T^d({\cal B}^d) \right \|_F^2$ in \eqref{eq:dual-Lagrangian-re} are relevant to ${\cal B}^d$. The gradient of $\frac{1}{2N_D} {\rm Tr}(\bm T^d({\cal B}^d))$ respect to ${\cal B}^d$ is given by
\begin{align}
\label{eq:derivative-2-nl-ap-p1}
\nabla_{{\cal B}^d(p_1,...,p_d)} \frac{1}{2N_D} {\rm Tr}(\bm T^d({\cal B}^d)) = \left\{\begin{array}{ll}{\frac{1}{2},} & {p_1=...=p_d=0,} \\ {0,} & {\text{otherwise}.}\end{array}\right.
\end{align}
The gradient of ${\rm Tr}\left[ (\bm{\bar\Theta} - \bm T^d({\cal B}^d))^H \bm U \right]$ respect to ${\cal B}^d$ is given by
\begin{align}
\label{eq:derivative-2-nl-ap-p2}
&\nabla_{{\cal B}^d(p_1,...,p_d)} \left\langle  \bm U , \bm{\bar\Theta} - \bm T^d({\cal B}^d) \right\rangle_{\Re} \nonumber \\
&= - \nabla_{{\cal B}^d(p_1,...,p_d)}  \sum_{p_1,...,p_d} \beta_{(p_1,...,p_d)} {\mathbb{P}}\left( (\bm T^d({\cal B}^d))^H \bm U \right) (p_1,...,p_d) \nonumber \\
&= - \beta_{(p_1,...,p_d)} {\mathbb{P}}\left( \bm U \right) (p_1,...,p_d).
\end{align}
Similarly, the gradient of $\frac{\rho}{2} \left \| \bm{\bar\Theta} - \bm T^d({\cal B}^d) \right \|_F^2$ respect to ${\cal B}^d$ is given by
\begin{align}
\label{eq:derivative-2-nl-ap-p3}
&\nabla_{{\cal B}^d(p_1,...,p_d)} \frac{\rho}{2} \left \| \bm{\bar\Theta} - \bm T^d({\cal B}^d) \right \|_F^2 \nonumber \\
&= \nabla_{{\cal B}^d(p_1,...,p_d)} \left\{ -\rho {\rm Tr}\left[  (\bm T^d({\cal B}^d))^H \bm{\bar\Theta} \right] + \frac{\rho}{2} {\rm Tr}\left[ (\bm T^d({\cal B}^d))^H \bm T^d({\cal B}^d) \right] \right\} \nonumber \\
&= \rho \beta_{(p_1,...,p_d)} \left( - {\mathbb{P}}\left( \bm{\bar\Theta} \right) (p_1,...,p_d) + {\cal B}^d(p_1,...,p_d) \right).
\end{align}
Summing up \eqref{eq:derivative-2-nl-ap-p1}, \eqref{eq:derivative-2-nl-ap-p2} and \eqref{eq:derivative-2-nl-ap-p3} yields \eqref{eq:derivative-2-nl}.

\subsection{Proof of Proposition~\ref{prop:dual}}

\subsubsection{Proof of \eqref{eq:lemma-dual-eq1}}

Note that the dual problem in \eqref{eq:dual-problem-1} is derived from the Lagrangian of \eqref{eq:AN-problem-1}, which is given by
\begin{align}
{{\xi}_1}({\cal X}, {\cal V}) =  \| {\cal X}\|_{{\mathbb{A}}({\mathbb{F}})} + \langle {\cal V}, {\cal Y} - {\cal P} \odot {\cal X} \rangle_{\Re}.
\end{align}
Moreover, the Lagrangian of \eqref{eq:SDP-problem-1} is given by
\begin{align}
\label{eq:dual-L}
{\bar\xi_1}(\bm x, {\cal B}^d, {t}, \bm \Theta, \bm{\widetilde U}) = &~ \frac{1}{2N_D} {\rm Tr}(\bm T^d({\cal B}^d)) + \frac{1}{2} {t} + \mathbb{I}_{\infty}(\bm \Theta \succeq 0) + \sum_{i=1}^d \mathbb{I}_{\infty}(\bm T_{g_i}^d({\cal B}^d) \succeq 0)   \nonumber \\
& + \left \langle \bm{U}, \left[ {\begin{array}{*{20}{c}} 
\bm \Theta - \left[ {\begin{array}{*{20}{c}}
	\bm T^d({\cal B}^d) & \bm x \\
	\bm x^H & {t}
	\end{array}} \right] \end{array}} \right]  \right \rangle_{\Re} + 
\left \langle \bm u, \bm y - \bm \Phi \bm x  \right \rangle_{\Re}.
\end{align}
Since \eqref{eq:AN-problem-1} and \eqref{eq:SDP-problem-1} are equivalent if ${\rm rank}(\bm T^d) < \min_i N_i $, $\bm u = {\rm vec}({\cal V})$ is the vectorized dual variable. Furthermore, by noting the Karush-Kuhn-Tucher (KKT) conditions~\cite{boyd2004convex}, the optimal value satisfies
\begin{align}
\label{eq:KKT-derivative-1}
&\nabla_{\bm x}{\bar\xi_1} = - \bm \Phi^H \bm{\widehat{u}} - 2\bm{\widehat{\bar{u}}} = 0,
\end{align}
which yields \eqref{eq:lemma-dual-eq1}.

\subsubsection{Proof of \eqref{eq:lemma-dual-eq2}}

We first find the conditions that the dual variable satisfies from the derivation of \eqref{eq:dual-problem-2}. Rewrite \eqref{eq:AN-problem-2} as 
\begin{align}
\label{eq:AN-problem-2-2}
\widehat{\cal X} = &~\arg \min_{{\cal X} } \frac{1}{2} \| {\mathcal{Y}} - {\cal Z} \|_F^2 + \lambda \| {\cal X}\|_{{\mathbb{A}}({\mathbb{F}})},\\
&~{\rm s.t.}~ {\cal Z} = {\mathcal{P}} \odot {\cal X}, \nonumber
\end{align}
whose Lagrangian is given by
\begin{align}
{{\xi}_2}({\cal X}, {\cal V},{\cal Z}) =&~ \lambda \| {\cal X}\|_{{\mathbb{A}}({\mathbb{F}})} + \frac{1}{2} \| {\cal Y} - {\cal Z} \|_F^2 + \langle {\cal V}, {\cal Z} - {\cal P} \odot {\cal X} \rangle_{\Re}.
\end{align}
Note that 
\begin{align}
&~\|{\cal V}\|_{{\mathbb{A}}({\mathbb{F}})}^* = \sup_{\|{\cal X}\|_{{\mathbb{A}}({\mathbb{F}})} \leq 1 } \langle  {\cal V},  {\cal P} \odot {\cal X} \rangle_{\Re} = \sup_{\|{\cal X}\|_{{\mathbb{A}}({\mathbb{F}})} = 1 } \langle  {\cal V},  {\cal P} \odot {\cal X} \rangle_{\Re} \nonumber \\
&~ \Rightarrow  \|{\cal V}\|_{{\mathbb{A}}({\mathbb{F}})}^* \|{\cal X}\|_{{\mathbb{A}}({\mathbb{F}})} \geq \langle  {\cal V},  {\cal P} \odot {\cal X} \rangle_{\Re},
\end{align}
which yields $\lambda \| {\cal X}\|_{{\mathbb{A}}({\mathbb{F}})} - \langle {\cal V}, {\cal P} \odot {\cal X} \rangle_{\Re} \geq (\lambda - \|{\cal V}\|_{{\mathbb{A}}({\mathbb{F}})}^*) \|{\cal X}\|_{{\mathbb{A}}({\mathbb{F}})}$. Hence the minimum of ${{\xi}_2}({\cal X}, {\cal V}, {\cal Z})$ with respect to ${\cal X}$ is $-\infty$ unless $\|{\cal V}\|_{{\mathbb{A}}({\mathbb{F}})}^* \leq \lambda$ holds. And if $\|{\cal V}\|_{{\mathbb{A}}({\mathbb{F}})}^* \leq \lambda$ holds then the minimum of ${{\xi}_2}({\cal X}, {\cal V}, {\cal Z})$ is at ${\cal X} = 0$.
Set ${\cal X} = 0$ and all that remains is to minimize
\begin{align}
\label{eq:remains-dual}
\frac{1}{2} \| {\cal Y} - {\cal Z} \|_F^2 + \langle {\cal V}, {\cal Z} \rangle_{\Re}
\end{align}
with respect to ${\cal Z}$. The function is convex with respect to ${\cal Z}$, hence we minimize \eqref{eq:remains-dual} by setting the gradient with respect to ${\cal Z}$ to zero
\begin{align}
\label{eq:gradientZ0}
- {\cal Y} + {\cal Z} + {\cal V} = 0,
\end{align}
which yields the dual problem in \eqref{eq:dual-problem-2}. Hence, from \eqref{eq:gradientZ0} we have the optimal dual variable $\widehat{\cal V}$ satisfies 
\begin{align}
\label{eq:dual-condition}
\widehat{\cal V} = {\cal Y} - {\mathcal{P}} \odot \widehat{\cal X} \Rightarrow \bm{\widehat\nu} = \bm y - \bm \Phi \bm{\widehat{x}}.
\end{align}

Moreover, the Lagrangians of \eqref{eq:SDP-problem-2} is given by
\begin{align}
\label{eq:dual-L-2}
{\bar{\xi}_2}(\bm x, {\cal B}^d, {t}, \bm \Theta, \bm U) = &~ \frac{1}{2} \| \bm y - \bm \Phi \bm x \|_2^2 + \frac{\lambda}{2N_D} {\rm Tr}(\bm T^d({\cal B}^d)) + \frac{\lambda}{2} {t} + \mathbb{I}_{\infty}(\bm \Theta \succeq 0) \nonumber \\
& + \sum_{i=1}^d \mathbb{I}_{\infty}(\bm T_{g_i}^d({\cal B}^d) \succeq 0) 
+ \left \langle \bm U, \bm \Theta - \left[ {\begin{array}{*{20}{c}}
	\bm T^d({\cal B}^d) & \bm x \\
	\bm x^H & {t}
	\end{array}} \right] \right \rangle_{\Re}.
\end{align}
The optimal value satisfies
\begin{align}
\label{eq:KKT-derivative-2}
\nabla_{\bm x}{\bar{\xi}_2} = \bm \Phi^H(\bm \Phi \bm{\widehat x} - \bm y) - 2\bm{\widehat{\bar{u}}} = 0.
\end{align}
Since \eqref{eq:AN-problem-2} and \eqref{eq:SDP-problem-2} are equivalent if ${\rm rank}(\bm T^d) < \min_i N_i $, combining \eqref{eq:dual-condition} and \eqref{eq:KKT-derivative-2} yields \eqref{eq:lemma-dual-eq2}. Then we complete the proof.

\bibliographystyle{IEEEtran}
\bibliography{database}

\end{document}